\documentclass[fleqn,usenatbib]{mnras}

\usepackage{newtxtext,newtxmath}
\usepackage{graphicx}

\usepackage[T1]{fontenc}

\DeclareRobustCommand{\VAN}[3]{#2}
\let\VANthebibliography\thebibliography
\def\thebibliography{\DeclareRobustCommand{\VAN}[3]{##3}\VANthebibliography}

\usepackage{graphicx}	
\usepackage{amsmath}	
\usepackage[english]{babel}
\usepackage{comment}

\title[Cavities]{Cavitation instability in unmagnetized relativistic pair shocks}

\author[Demidov et al.]{
I. Demidov\thanks{E-mail: dvsmallville@gmail.com}, Y. Lyubarsky,
U. Keshet
\\
Physics Department, Ben-Gurion University, PO Box 653, Beer-Sheva 84105, Israel\\
}

\date{Accepted XXX. Received YYY; in original form ZZZ}

\pubyear{2025}

\begin{document}
\label{firstpage}
\pagerange{\pageref{firstpage}--\pageref{lastpage}}
\maketitle

\begin{abstract}
We investigate the formation of plasma cavities in unmagnetized relativistic pair shocks and demonstrate that these cavities emerge naturally as a nonlinear consequence of asymmetric Weibel instability. We provide an intuitive physical picture and a minimal fluid model that captures the essential features of this cavitation instability and compare them with PIC results. This mechanism may provide the missing link between kinetic Weibel turbulence and macroscopic magnetic fields in astrophysical shocks.

\end{abstract}




\section{Introduction}

Understanding the
fundamental problem of magnetic field generation in collisionless unmagnetized shocks
is crucial for explaining the observed radiative signatures of GRB afterglows.
The afterglow is well described by synchrotron models (e.g. \citealt{Waxman1997}; \citealt{Sari1998}; \citealt{GruzinovWaxman1999}).
However, the GRB external blast wave propagates into plasma of very low 
magnetization, with the magnetization parameter $\sigma\sim 10^{-9}-10^{-5}$, while synchrotron radiation models require the formation of much stronger magnetic fields at large distances $\sim 10^{10}d_e$, where $d_e$ is the plasma skin-depth.

In this unmagnetized environment, the main mechanism for shock formation is the Weibel instability (current filamentation), which can generate near equipartition magnetic fields at kinetic scales (e.g. \citealt{MedvedevLoeb1999}; \citealt{Brainerd2000}; \citealt{Lemoine2019}). However, these magnetic fields decay very rapidly behind the shock, on a scale of the order of several hundred plasma skin-depths (\citealt{Gruzinov2001}; \citealt{Chang2008}; \citealt{Lemoine2015}), and cannot sustain the observed emission unless structure somehow grows to large scales through some self-similar organization  \citep{Katz2007}.
This apparent contradiction constitutes the GRB afterglow magnetization problem.

Recent particle-in-cell (PIC) simulations of relativistic shocks offer a possible way to resolve this paradox. In pair plasma (\citealt{Keshet2009}; \citealt{Groselj2024}), in pair-loaded electron-ion plasma (\citealt{Groselj2022}), and in electron-ion plasma (\citealt{Naseri2018}; \citealt{Peterson2021, Peterson2022}; \citealt{Bresci2022}), at sufficiently long evolution times, the Weibel-mediated shock precursor develops nonlinear structures with an enhanced magnetic field. While such structures seem to be a generic feature, in this work, we focus on the case of shocks in pair plasma.

\begin{figure*}
\begin{minipage}{0.95\textwidth}
    \centering
    \includegraphics[width=\linewidth]{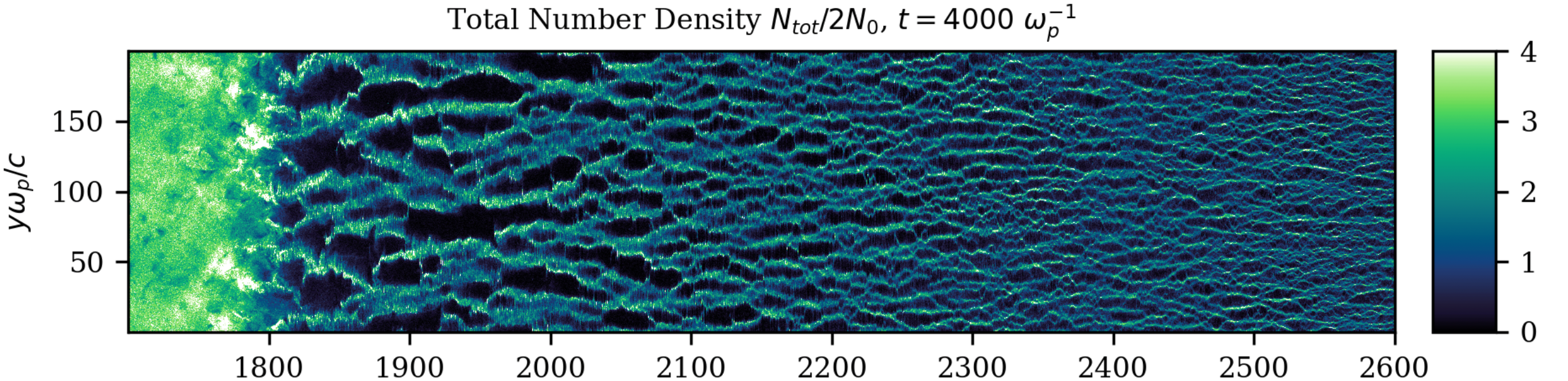}
    \label{fig:shockD}
\end{minipage}%
\hfill
\begin{minipage}{0.96\textwidth}
    \centering
    \includegraphics[width=\linewidth]{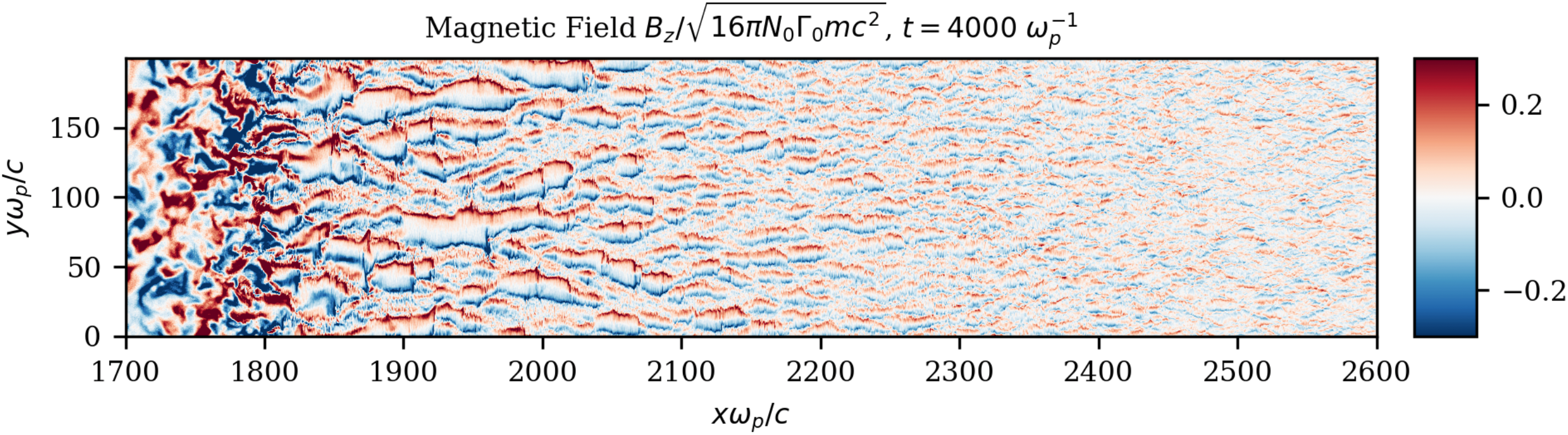}
    \caption{Distribution of total plasma density and magnetic field $B_z$ in a PIC simulation of a relativistic shock propagating into an unmagnetized  pair plasma. The shock precursor exhibits a cellular pattern, with plasma collecting in dense walls where the $B$-field is the strongest. The simulation is in the downstream frame, where the upstream plasma Lorentz factor is $\Gamma_0=10$. The shock transition at the presented time, $4000\omega_p^{-1}$, is located at $\sim 1800(c/\omega_p)$ and moves to the right with velocity 
    $v_\text{sh}\sim c/2$. All simulation parameters are listed in Appendix~\ref{sec:numerical}.
    }
    \label{fig:shockM}
\end{minipage}
\end{figure*}

The upstream precursor undergoes highly nonlinear filamentation, producing narrow filaments and low-density cavities. The overdense walls around the cavities form a cellular pattern in the shock precursor, as shown in the upper panel of Fig.~\ref{fig:shockM}. The strongest magnetic field is observed within these narrow plasma walls, while the interiors of the cavities remain nearly field-free (bottom panel of Fig.~\ref{fig:shockM}). This picture is in contrast to the linear Weibel stage, where 
the plasma accumulates at the nodes of the magnetic field.
As plasma approaches the shock, some cavities expand in size. When these large cavities cross the shock front, they increase the coherence length of the magnetic field in the downstream. 
If this length exceeds several times the average Larmor radius of particles, the downstream plasma becomes magnetized and "holds"\, the magnetic field, slowing down its decay. At the same time, increasing the characteristic scale of magnetic field fluctuations makes it possible to accelerate particles to higher energies (e.g. \citealt{Keshet2009}; \citealt{Sironi2013}; \citealt{RevilleBell2014}).
According to PIC simulations, in the downstream, most of the magnetic energy is concentrated in intermittent structures, with localized patches reaching near-equipartition fields, while the volume-averaged field remains much lower (e.g. \citealt{Groselj2024}). Such intermittency implies that most of the synchrotron emission is produced in the strong-field regions, despite their small filling fraction. This could resolve the GRB afterglow magnetization paradox if these regions with a significant magnetic field were to persist over long distances behind the external shock.





One of the first open questions is how the shock precursor develops its cellular structure, with plasma cavities separated by magnetized walls, in contrast to the linear Weibel picture. In this work, we focus on this problem, presenting a qualitative physical picture together with a minimal mathematical model and supporting numerical simulations of cavity formation in unmagnetized relativistic pair plasma shocks.  Using homogeneous PIC simulations, backed by non-magnetized shock simulations, we clarify how cavities emerge self-consistently from the nonlinear evolution of Weibel filaments.
This paper is organized as follows. In Section~\ref{sec:2} we describe the structure of cavities that arise in homogeneous PIC simulations. In Section~\ref{cavform}, we give an intuitive explanation of the cavity formation process with qualitative estimations. In Section~\ref{sec:3} we present a minimal self-consistent mathematical model for the cavitation process and compare its solutions with homogeneous PIC simulations. Finally, in Section~\ref{sec:4} we discuss the connection with Weibel shocks and provide the discussion and summary of our results. Appendix~\ref{sec:boosted} provides an analysis of the cavity formation in the co-moving frame of the denser beam, corresponding approximately to the upstream fluid frame in a shock.

\section{Cavity formation in homogeneous simulations}\label{sec:2}

First, let us describe the structure of these cavities. To do this, we use a spatially homogeneous PIC setup with counter-streaming beams chosen to mimic the plasma conditions in a shock precursor. This configuration isolates the essential microphysical mechanism from the global shock dynamics, enabling us to track the formation of cavities under controlled conditions. We find that cavities indeed emerge self-consistently from the non-linear evolution of Weibel filaments.

PIC simulations are performed with the SMILEI code \citep{Derouillat2018} in 2D geometry in the $x$-$y$ plane, with flow in the $x$-direction. Homogeneous simulation is periodic in both directions. To suppress longitudinal modes and retain purely transverse Weibel physics, the longitudinal (i.e., $x$-direction) box size is chosen shorter than the shortest unstable parallel wavelength. Full numerical and physical parameters are listed in Appendix~\ref{sec:numerical}.

In our reference run, mostly corresponding to the early evolution of the shock, a cold background pair plasma (subscript $_\text{p}$) drifts against the $x$-axis with a Lorentz factor
$\Gamma_{\rm p} \equiv\Gamma_0= 10$, while a dilute, hot beam (subscript $_\text{b}$) propagates in the opposite
direction with $\Gamma_{\rm b} = 3$.
The density ratio is $N_{\rm b} / N_{\rm p} = 6\times10^{-3}$; proper temperatures are $T_{\rm p}=10^{-3}mc^{2}$ and $T_{\rm b}=10\,mc^{2}$ expressed in units of the electron rest energy. The box size is $L_x\times L_y=8\times32\,(c/\omega_{\rm p})^2$, where $\omega_{\rm p}^{2}=8\pi N_{\rm p}/(\Gamma_{\rm p} m)$ is the relativistic plasma frequency with respect to the total background plasma density. This configuration is intrinsically asymmetric, as a dense plasma interacts with a dilute beam, and this asymmetry is crucial for the subsequent cavity formation.
    \begin{figure}
    \begin{minipage}{0.48\textwidth}
        \centering
        \includegraphics[width=\linewidth]{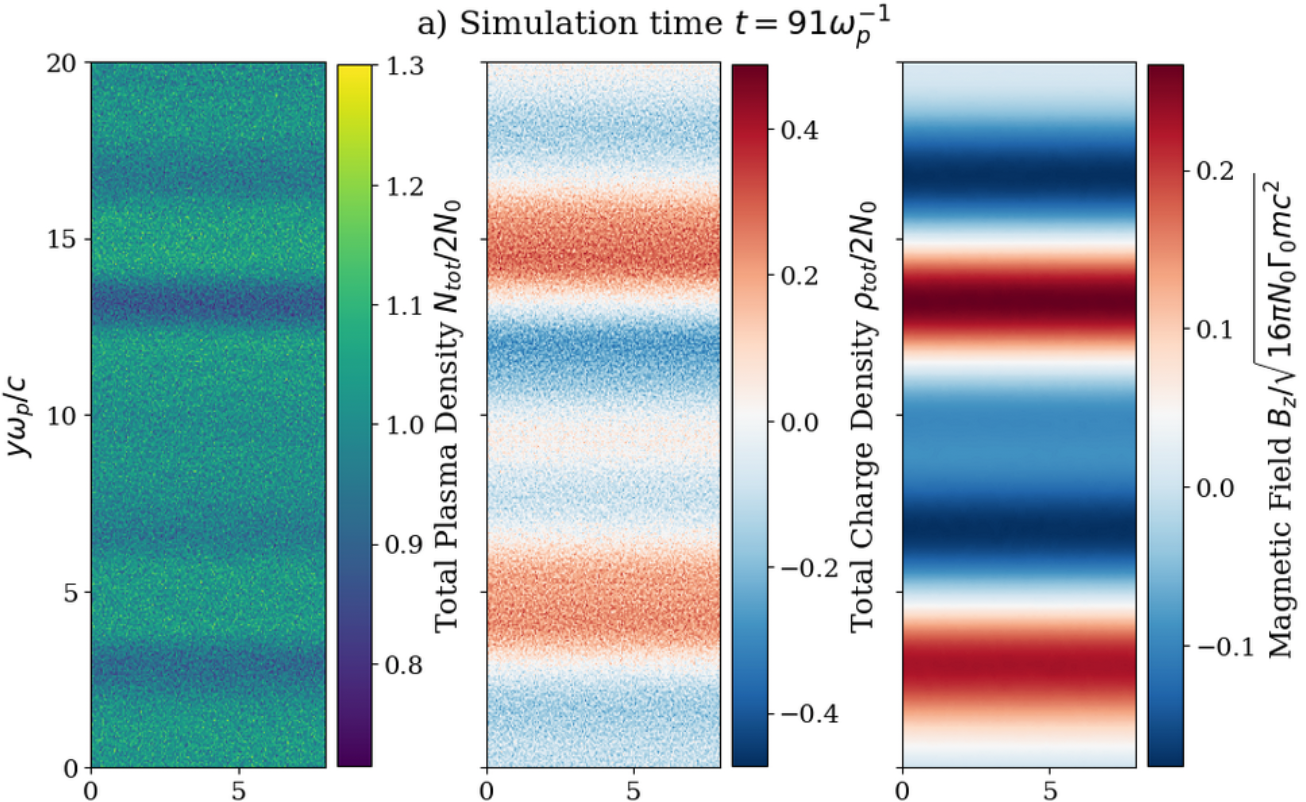}
        \label{fig:lab1}
    \end{minipage}%
    \hfill
    \begin{minipage}{0.48\textwidth}
        \centering
        \includegraphics[width=\linewidth]{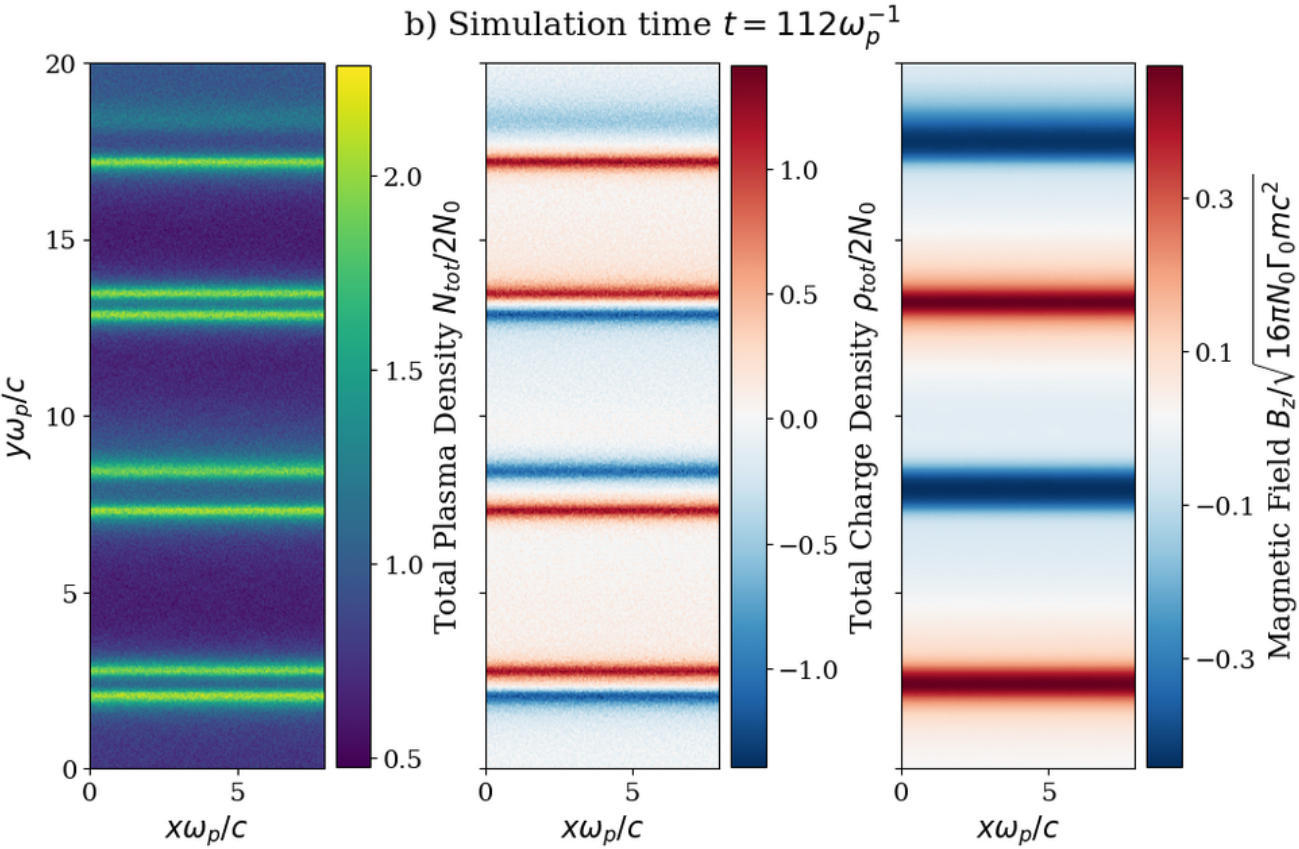}
        \caption{Reference run distributions of total plasma density, total charge density, and magnetic field $B_z$ in the lab frame $S$ at different simulation times: a) just before cavity formation; b) when cavities are formed as magnetic fields and plasma gather into thin walls at the edges of the filament. The figure shows only the part of the simulation box, $y<20(c/\omega_p)$, where the cavity walls are the strongest.}
        \label{fig:lab2}
    \end{minipage}
    \end{figure}

The simulation results of this reference run are shown in Fig.~\ref{fig:lab2}. Initially, there is a regular Weibel instability. During this stage, the denser plasma beam experiences only weak density modulation, forming, together with the dilute beam, filaments with a radius $R$ of the order of several plasma skin depths. The density and magnetic field modulations differ in phase by $90^\circ$, such that the plasma accumulates at the magnetic field nodes.

The dilute beam reaches saturation, i.e. almost complete $e^-/e^+$ separation, at $t\sim 90\omega_p^{-1}$. After beam saturation, denser plasma begins to migrate from the center of the filament to its edges, where the magnetic field is maximal. 
The cavity is then formed in about $\sim 10\omega_p^{-1}$. The transverse motion of the plasma compresses the magnetic field, forming a kind of "magnetic sandwich": focused and compressed electron and positron walls of adjacent filaments, between which there is an enhanced magnetic field. At the same time, the magnetic field inside the filaments decreases sharply.

The decrease in magnetic field inside the cavity is evident when approximating the walls of a cavity as two planes of parallel currents. Almost all the current is concentrated in these charged walls owing to their large density. According to the superposition rule, the filament-induced magnetic field between the planes is zero, and is doubled outside. The remaining magnetic field inside the cavity is created by the beam, i.e., returning particle current and a small fraction of the remaining background plasma. The walls of adjacent cavities cannot touch each other because the separating plasma is magnetized, and the magnetic pressure pushes them apart. On the other hand, the walls of adjacent cavities are attracted to each other by the electrical force (like the plates of a capacitor).

The same behavior (amplification at the edges of the cavities and significant weakening in its center) is observed for the transverse electric field, since $|E_y|\simeq|B_z|$ as shown later. The longitudinal electric field, $E_x$, inside the filaments decreases and changes sign during cavity formation (not shown in Fig.~\ref{fig:lab2}, but see Section~\ref{sec:3}).

\begin{figure}
\begin{minipage}{0.48\textwidth}
    \centering
    \includegraphics[width=\linewidth]{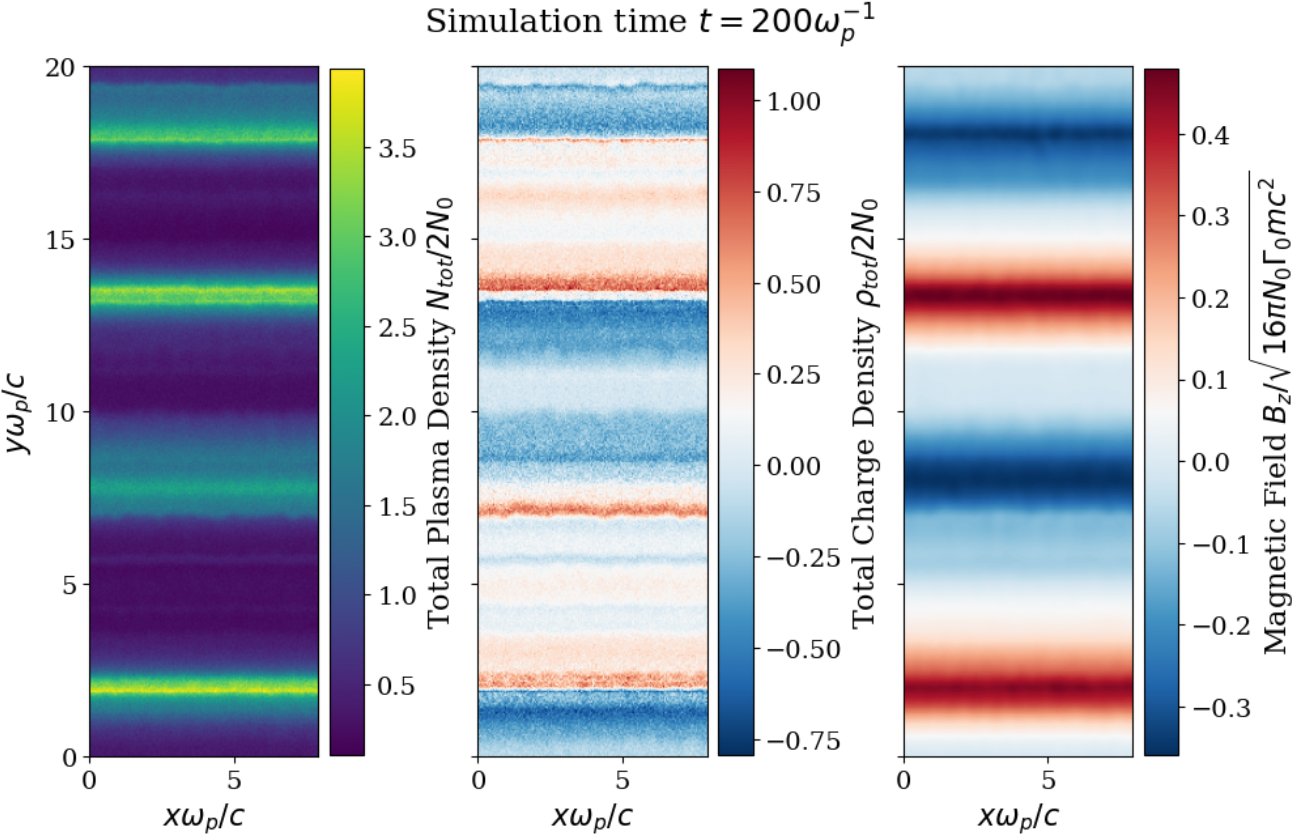}
    \label{fig:lab3}
\end{minipage}%
\hfill
\begin{minipage}{0.48\textwidth}
    \centering
    \includegraphics[width=\linewidth]{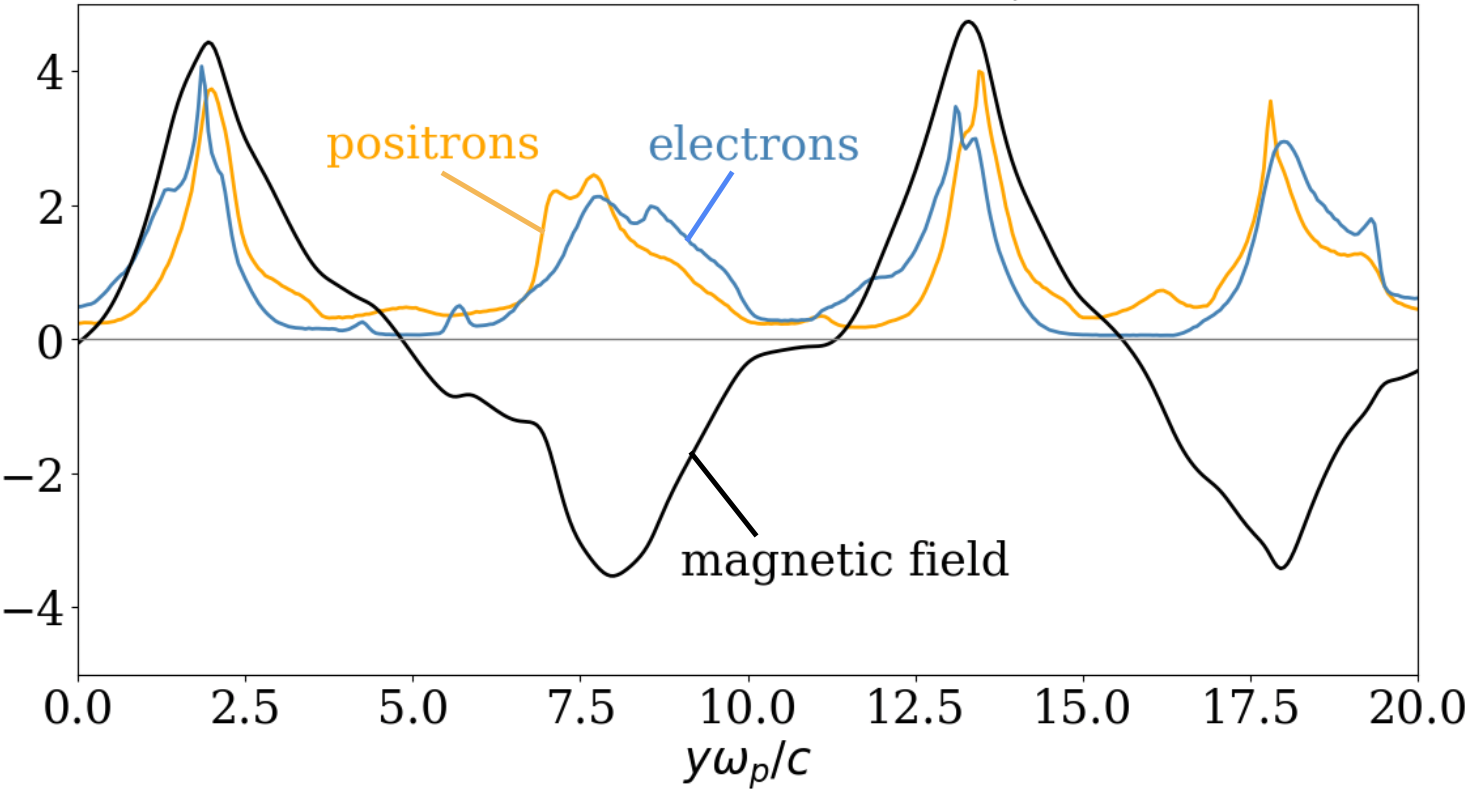}
    \caption{Reference run distributions of total plasma density, total charge density, and magnetic field $B_z$ in the lab frame $S$ at $t=200\omega_p^{-1}$. Top panel -- same as Fig.~\ref{fig:lab2}; bottom panel -- profiles of $N_+/N_0$ (positrons), $N_-/N_0$ (electrons), and $B_z/\sqrt{16\pi N_0\Gamma_0 mc^2}$ averaged over the $x$-coordinate (for clarity, the magnetic field is scaled by a factor of $\Gamma_0$).}
    \label{fig:lab4}
\end{minipage}
\end{figure}

The beam particles propagate between the plasma walls and fill the cavities.
The emergence of cavities is accompanied by background plasma heating, which increases the wall thickness. Nevertheless, despite this broadening, the walls remain coherent and persist over long timescales in the simulations; therefore, the cellular pattern of shock precursor and magnetized walls can be maintained. This structure of "magnetic sandwich" is not permanent. As the plasma compresses and heats, its walls diffuse into the region of strong field (see Fig.~\ref{fig:lab4}). Electrons and positrons undergo collective oscillations in the walls, resulting in oscillations of both the magnetic field extrema and the wall thicknesses.




\section{Cavities formation in the lab-frame $S$}\label{cavform}


\subsection{Qualitative description}\label{qual}

Two oppositely directed relativistic plasma beams are unstable to the Weibel/current filamentation instability (\citealt{Weibel1959}, \citealt{Fried1959}). Small, random magnetic fluctuations seed tiny sideways deflections of the charges, which slightly modulate the current density. By Ampère's law, these currents amplify magnetic fluctuations, providing positive feedback. The flow thus breaks into narrow current filaments with a characteristic radius set by the plasma skin depth. The magnetic field focuses oppositely charged particles from the dilute beam and background plasma flow into the same current filaments.
Since the number densities of the flows are vastly different, the filaments are charged, and a transverse electrostatic field $E_y$ builds up between neighboring filaments. If only $B_z$ were present, both the beam and the background plasma would be focused into filaments at the same rate, leading to a large excess of charge due to $N_\text{p}\gg N_\text{b}$. Thus, the electrostatic field is an inherent property of the asymmetric Weibel instability.
 
\begin{figure}
  \centering
  \includegraphics[width=\columnwidth]{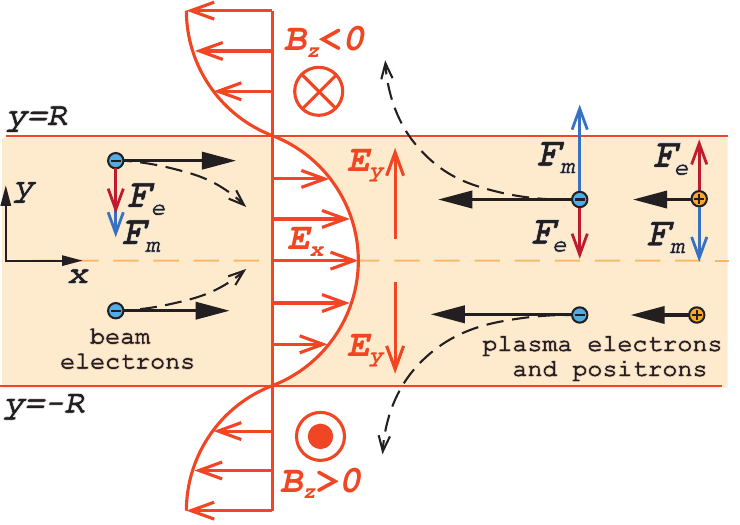}
  \caption{Sketch of a plasma positron-dominated Weibel filament. Black arrows show the longitudinal component of velocities; $F_e$ and $F_m$ denote the transverse electric and magnetic forces, respectively.
  }
\label{fil}
\end{figure}

As an example, we consider in Fig.~\ref{fil} a filament centered at $y=0$, in which plasma positrons together with beam electrons are focused, whereas plasma electrons and beam positrons are pushed out. 
Given that $N_{\rm p}\gg N_{\rm b}$, we refer to this filament as positron-dominated. 

Since the particle densities enter with opposite signs in the charge density, $\rho\simeq e(N_{\rm p}-N_{\rm b})$, but with the same sign in the current, $j\simeq e(N_{\rm p}+N_{\rm b})c$, the magnetic field exceeds the electric field during the linear Weibel stage. The dominance of the magnetic field causes the filament pattern grow. However, since the filament is positively charged, the electrostatic field $E_y$ slows down the focusing of the plasma positrons into the filament while helping the beam electrons to focus. If the beam is very dilute, the difference $B^2_z-E^2_y$ should be very small;
for the plasma particles, the magnetic and electric forces nearly cancel, whereas for the dilute beam particles, they act in the same direction, resulting in an overall focusing roughly twice as strong (see Fig.~\ref{fil}).

As $B_z$ grows during the linear Weibel stage, it induces by Faraday's law a longitudinal electric field $E_x$ inside the filament. Since the particles are initially essentially unmagnetized, $E_x$ directly changes their longitudinal velocities. By Lenz’s law, $E_x$ is directed to decelerate the plasma positrons, which carry the dominant current. Hence, the plasma electrons must be accelerated by this field. Therefore, inside the filament, the plasma electrons acquire a slightly larger average longitudinal velocity than the positrons, and the magnetic force, $F_m=|(e/c) B_z v_x|$, acting on the electrons becomes correspondingly stronger. 

As the instability develops, two scenarios are possible: (i) If the dilute beam reaches saturation, $N_{\rm b}$ stops increasing, whereas $N_{\rm p}$ continues to grow and the plasma positrons keep being slowed down by $E_x$. As a result, the charge density increases faster than the current density, since the latter also depends on the positron velocity, which decreases with time. Consequently, the electrostatic field, $E_y$, grows more rapidly than the magnetic field. Therefore, near the filament center, where the positrons are slowed down the most, the transverse electric and magnetic forces acting on them eventually come into balance, $F_e=F_m$. However, the magnetic force on the expelling plasma electrons is larger because they are accelerated by $E_x$ and move faster. Thus, electrons keep leaving the filament center and the positive charge continues to grow, as well as the electrostatic field $E_y$. Eventually, the force balance on the positrons breaks down, and the resulting transverse force acting on them becomes expelling. (ii) If the plasma positrons are decelerated strongly enough already during the linear Weibel phase, their reduced longitudinal velocity may weaken the magnetic force to the point where the electric force dominates before the dilute beam fully saturates. Indeed, when the beam density is very low, the magnetic field has only a modest advantage over the electric field; thus, once this field is multiplied by the reduced positron velocity, the magnetic force can become smaller than the electric force. The equilibrium of positrons becomes impossible, and they are pushed out of the filament center. Since the number density of plasma positrons in the filament dominates, it leads to cavity formation. Formally, both scenarios follow from a single inequality (see equation~\ref{eq:inequality} in the next subsection); the two outcomes arise from different parameter ranges. 

Note that for the dilute beam particles, the electric and magnetic forces act in the same direction; therefore, the beam cannot form cavities. Obviously, this mechanism should also work in the 3D case, where electric and magnetic forces are directed radially with respect to a filament.

\subsection{Back-of-the-envelope estimations}

It is convenient to introduce a velocity $V_{w}$ that connects fields $E_y$ and $B_z$. Our system is magnetically dominated, $B_{z}^{2}-E_{y}^{2}>0$, therefore, one can define the Weibel frame $W$ in which the charge of the filament is zero, and there is no electrostatic field, $E_{y|w}=0$ (e.g. \citealt{Pelletier2019}). Then the Lorentz transformation between $W$ and $S$ gives
\begin{equation}\label{eq:Ey}
  E_{y} = \frac{V_{w x}}{c}\, B_{z},
\end{equation}
where $V_{wx}$ denotes the $x$-component of the velocity of frame $W$ relative to~$S$. Physically, this velocity represents the $\mathbf{E}\times\mathbf{B}$ drift of the plasma. In the linear and homogeneous Weibel theory with a single transverse wavenumber $\mathbf{k}\perp\hat{\mathbf{x}}$, fields $E_y$ and $B_z$ share the same $y$-dependence (up to phase), so the ratio $E_y/B_z$ is independent of $y$. This implies a unique boost along $x$ for the given mode $k$.

\begin{figure}
  \centering
  \includegraphics[width=0.75\columnwidth]{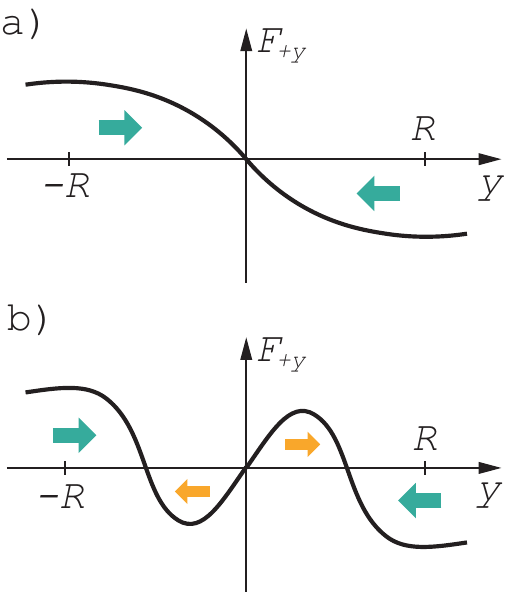}
  \caption{Sketch of total Lorentz force acting on positrons in the plasma positron filament: a) during the linear Weibel stage; b) during cavity formation. 
  The arrows show the direction of positron acceleration under the action of force $F_{+y}$.
  }
\label{force}
\end{figure}

By using the Weibel frame velocity, the transverse Lorentz force that acts on positrons can be written in a compact form,
\begin{equation}
  F_{+y} = e\!\left(E_{y} - \frac{v_{+x}}{c} B_{z}\right)
          = \frac{e}{c} \bigl(V_{w x} - v_{+x}\bigr) B_{z},
\end{equation}
where $v_{+x}$ is the $x$-component of the positron velocity. Initially, $|v_{+x}| > |V_{w x}|$ across the entire cross section of the filament, such that the force $F_{+y}$ is focusing and leads to filamentation.  As $E_{x}$ decelerates the positrons, there appears a region inside the filament where
\begin{equation}\label{eq:inequalityV}
  |v_{+x}| < |V_{w x}|,
\end{equation}
or equivalently
\begin{equation}\label{eq:inequality}
    \Gamma_+<\Gamma_w.
\end{equation}
Here $\Gamma_+$ and $\Gamma_w$ denote the Lorentz factors of the plasma positrons and of the Weibel frame $W$, respectively. In this region the electrostatic component wins and the force becomes defocusing: positrons migrate toward the edges, effectively creating a cavity (see Figure~\ref{force}). The region where the force $F_{+y}$ is defocusing increases with time, since at greater distances from the filament center (where the electric field $E_x$ is weaker) the positrons need more time to slow down sufficiently. The resulting inequality~(\ref{eq:inequalityV}) also means that in the Weibel frame, positrons reverse the direction of motion; then the magnetic force acting on them also changes sign and becomes defocusing.

Since the velocities are relativistic, it is more convenient to work with Lorentz factors. Since everything is approximated here as uniform along $x$, the canonical momentum is conserved for each particle. Therefore, in the filament center, $y=0$, we have
\begin{equation}
    \Gamma_{+}mv_{+x}+\frac{e}{ck}B_{\rm max}=\Gamma_0mv_{0x},
\end{equation}
where $B_{\rm max}$ is the amplitude of the magnetic field, and $v_{0x}$ is the initial positron velocity before formation of Weibel filaments.
Assuming $\Gamma_+\gg1$ and $\Gamma_0\gg1$, we obtain
\begin{equation}
    \Gamma_+\simeq\Gamma_0-\frac{eB_{\rm max}}{mc^2k}.
\end{equation}
Solving the inequality $\Gamma_+<\Gamma_w$ with respect to the amplitude of the magnetic field, we obtain
\begin{equation}\label{can}
    B_{\rm max}>\frac{kmc^2}{e}(\Gamma_0-\Gamma_w).
\end{equation}
This inequality also means that the positron deceleration time $\Delta t\sim mc(\Gamma_0-\Gamma_w)/eE_x^{\rm max}$, where $E_x^{\rm max}\simeq\gamma_w B_{\rm max}/ck$ from the Faraday equation, is smaller than the instability time $\sim \gamma_w^{-1}$. 
After the saturation, the transverse electric field $|E_y|$ grows faster than $|B_z|$, and consequently $\Gamma_w$ increases until inequality~(\ref{can}) can be satisfied even without significant slowing down of positrons.

\begin{figure}
\centering
\includegraphics[width=\columnwidth]{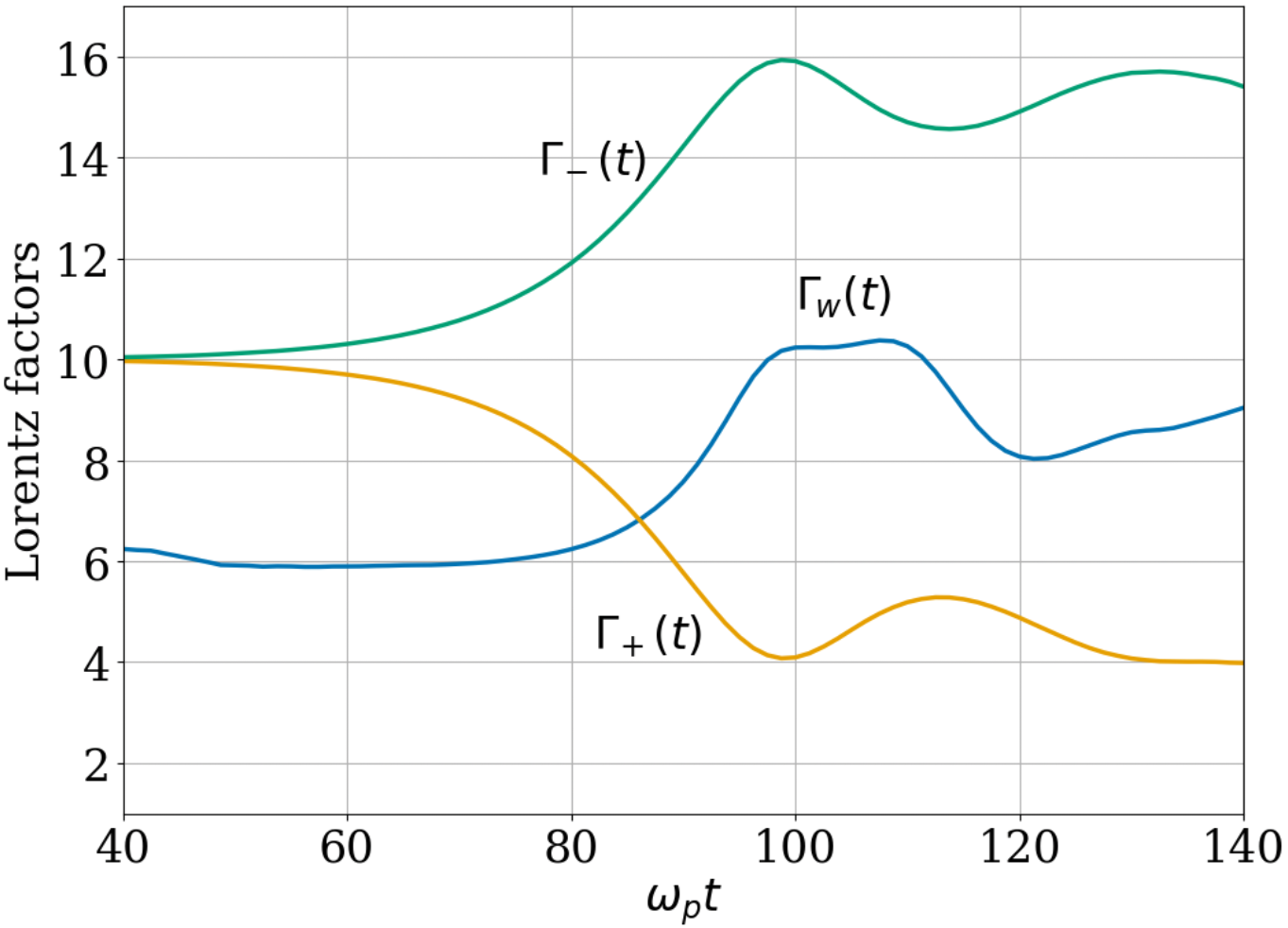}
\caption{Reference homogeneous run evolution of the Lorentz factors of plasma positrons $\Gamma_+$ and electrons $\Gamma_-$ at the positron-dominated filament center, as well as the Lorentz factor of the Weibel frame $\Gamma_w$.}
\label{gammaW}
\end{figure}

The cavity formation criterion, $\Gamma_+<\Gamma_w$, can be verified by PIC simulations. We tracked the evolution of the Lorentz factor of the Weibel frame. To avoid divisions by zero, we estimated this velocity as $V_w=c(\langle E_y^2\rangle/\langle B_z^2\rangle)^{1/2}$, where $\langle...\rangle$ denotes averaging over the simulation box (e.g. \citealt{Bresci2022}). 
Also, we computed the Lorentz factors of electrons and positrons at the positron-dominated filament center. Fig.~\ref{gammaW} shows that, right before the cavity is formed, the positrons indeed move more slowly than the Weibel frame. It can also be seen that the inequality $\Gamma_+<\Gamma_w$ then persists over time; therefore, the positrons do not return to the center of the filament.

Since the Weibel frame velocity $V_{wx}$ (but not the Lorentz factor $\Gamma_w$) changes insignificantly, 
differentiating the relation $E_y=(V_{wx}/c)B_z$ with respect to time $t$, we approximate $V_{wx}$ as constant. Substituting $\partial E_y/\partial t=-4\pi j_y$ and $\partial B_z/\partial t=c\,\partial E_x/\partial y$ from Maxwell's equations, and using $V_{wx}\simeq -c$, we obtain the relation
\begin{equation}\label{elec}
\frac{\partial E_x}{\partial y}\simeq\frac{4\pi}{c}j_y.
\end{equation}
This result can be explained as follows. The growing magnetic field creates, on the one hand, a vortex electric field $E_x$ according to Faraday's law\footnote {In neighboring filaments, $E_x$ is directed in different directions and thus forms vortices $E_x$ strongly elongated along the filaments.}, and on the other hand, an increasing transverse current $j_y$ due to the different deviation of charges of opposite signs. The current $j_y$ accumulates charge $\rho$ inside the filaments and increases the electrostatic field $E_y$ between them. In order for the fields $E_y$ and $B_z$ to be approximately equal and grow at the same rate, the transverse current must be proportional to the $z$-component of the vortex $E_x$, which gives ~(\ref{elec}).

Using the charge conservation law in the $S$ frame, we obtain
\begin{equation}\label{rhoE}
\frac{\partial \rho}{\partial t}\approx-\frac{c}{4\pi}\frac{\partial^2E_x}{\partial y^2}.
\end{equation}
Thus, regardless of the equations of particle motion, the sign of the second derivative $\partial^2 E_x/\partial y^2$ at a given point determines the local evolution of the charge density in the filament.
If $\partial^2 E_x/\partial y^2 < 0$ at a given point, positive charge accumulates there; if $\partial^2 E_x/\partial y^2 > 0$, negative charge accumulates.
Therefore, if regions with different signs of $\partial^2 E_x/\partial y^2$ exist within the filament, charge separation develops inside it.

The same mechanism operates symmetrically for plasma electron-dominated filaments, with the roles of the charges exchanged. In full shock simulations, the global
precursor provides a continual supply of returning particles that drives bursts of cavity
production \citep{Keshet2009, Groselj2024}.
Our homogeneous setup proves that the underlying instability cavity forming is intrinsic to the beam–plasma system itself. The formation of dense plasma walls does not require beam inhomogeneity or accelerated particles, which are present in shocks. However, these factors play a role in the expansion of these cavities. A similar conclusion was reached in \cite{Peterson2021, Peterson2022} and \cite{Bresci2022} for an electron–ion plasma; however, in that case the mechanism of cavity formation is different (see also \citealt{Califano2002}).

For a deeper understanding, we also consider the formation of cavities in the inertial reference frame $S'$, defined as co-moving initially with the background plasma frame. In this reference frame, the problem becomes inhomogeneous in the $x$-direction owing to the relativity of simultaneity; however, it can still be reduced to one spatial dimension. In $S'$, the electric and magnetic forces essentially swap roles: the electric force accumulates the background plasma into filaments, and the magnetic force repels it. As soon as the plasma inside filaments is accelerated by the longitudinal field $E_x$ to a certain velocity, the magnetic force wins and leads to the formation of a cavity. 
The details are given in Appendix~\ref{sec:boosted}.

\section{Toy model}\label{sec:3}

In this section, we consider the model sketched in Fig.~\ref{fil}. The dilute beam is assumed to be saturated and separated into electrons and positrons; henceforth, it is treated as dynamically passive, supplying constant current and charge over time. The background plasma has a weak transverse modulation, and its subsequent evolution is analyzed in a two-fluid description. This model can only describe cavities that form after beam saturation, even if the background plasma deceleration is insignificant (see Case 1 below). It cannot describe the other case, where the beam is not yet saturated, but the plasma has already slowed down enough to form cavities (Case 2). In this case, it is necessary to consider the self-consistent non-linear dynamics of the beam, for which a kinetic theory is needed. However, the simple model discussed below can give qualitatively correct results even in this case.

Let us write out the complete system of equations in the $S$ frame. The system is homogeneous along the $x$-axis; therefore, Maxwell's equations are
\begin{equation}\label{maxwell}
\begin{split}
    &\frac{\partial E_y}{\partial y}=4\pi\rho, \\
    &\frac{\partial E_x}{\partial y}=\frac{1}{c}\frac{\partial B_z}{\partial t}, \\
    &\frac{\partial B_z}{\partial y}=\frac{4\pi}{c}j_{x}+\frac{1}{c}\frac{\partial E_x}{\partial t}, \\
    &\frac{1}{c}\frac{\partial E_y}{\partial t}=-\frac{4\pi}{c}j_{y}.
\end{split}
\end{equation}
where charge and current densities are
\begin{equation}
    \rho=\sum_sq_s N_s, \quad j_{x}=\sum_s q_s N_s v_{sx}, \quad j_{y}=\sum_s q_s N_s v_{sy},
\end{equation}
and the $s$ index numerates all plasma species. These equations must be supplemented by the continuity equation:
\begin{equation}
    \frac{\partial N_s}{\partial t}+\frac{\partial}{\partial y}N_s v_{sy}=0.
\end{equation}
For simplicity, the background plasma is approximated as cold. Thus, to describe the motion of plasma, we can use the Euler equation,
\begin{equation}
    \frac{\partial \mathbf{v}_s}{\partial t}+v_{sy}\frac{\partial \mathbf{v}_s}{\partial y}=\frac{q_s}{\Gamma_s m}\left[\mathbf{E}+\frac{1}{c}(\mathbf{v}_s\times\mathbf{B})-\frac{\mathbf{v}_s(\mathbf{v}_s\cdot\mathbf{E})}{c^2}\right].
\end{equation}
The last term in the R.H.S. is related to the fact that we write the equation for $\mathbf{v}_s$ rather than for $\Gamma_s\mathbf{v}_s$ and the energy equation, $\partial\Gamma_s/\partial t+\mathbf{v}_s\cdot\nabla\Gamma_s=q_s(\mathbf{v}_s\cdot \mathbf{E})/mc^2$, was used. According to PIC simulation, the formation of walls is accompanied by heating of the plasma. Therefore, our cold model can only describe the formation of the cavity walls, but not their further evolution.

The number density of the beam is described phenomenologically. In our model, the beam is close to saturation, and its density does not change significantly over time; therefore, its average velocities will be considered as constants. At high beam temperatures, spatial inhomogeneity along the $y$-direction becomes pronounced. High $T_{\rm b}$ suppresses the Weibel growth rate, so filaments evolve much more slowly than the cavity-formation time. Once a some cavity is formed, it expands significantly, since neighboring filaments evolve too slowly to turn into cavities as quickly. Consequently, cavities are formed randomly within some of the strongest filaments and grow well beyond the initial filament radius. In this regime our assumption of time-independent beam behavior breaks down. The underlying formation mechanism of cavities remains the same; only the timescales and spatial uniformity change at high $T_{\rm b}$. Thus, we do not treat this high-temperature regime here and will analyze it elsewhere.

For the numerical solution, it is convenient to switch to dimensionless quantities.
For comparison with PIC simulations, let us use the normalization of the SMILEI code \citep{Derouillat2018}:
\begin{equation}
\begin{split}
&\mathbf{v}=c\mathbf{u}, \,\,\, t=\frac{1}{\omega_p}\tau, \,\,\, y=\frac{c}{\omega_p}\eta, \,\,\, \\
&N=\frac{m\omega_p^2}{4\pi e^2}n, \,\,\, \{E_{x,y},B_z\}=\frac{mc\omega_p}{e}\{\varepsilon_{x,y},b_z\}.
\end{split}
\end{equation}
where $\omega_p^2=8\pi N_0e^2/\Gamma_0 m$ is the relativistic plasma frequency in the laboratory reference frame $S$.

We again consider a plasma positron filament centered at $\eta=0$ and prescribe initial conditions motivated by linear filamentation as follows:
the particle densities are set as $n_{\pm}=n_0(1\pm\alpha\cos\eta')$; Lorentz factors as $\Gamma_{\pm}=\Gamma_{0}(1\pm\xi\cos\eta')$, and the transverse velocity as $u_{\pm y}=\mp\zeta\sin\eta'$. Here we introduce the shorthand $\eta'=(kc/\omega_p)\eta$ and $\pm$ denotes plasma species. The parameters $n_0$, $n_{b0}$, $\alpha$, $\xi$, and $\zeta$ are taken directly from the PIC simulations, at the moment when $\Gamma_+\approx\Gamma_w$ at $\eta=0$ (which gives $\xi\approx\Gamma_w/\Gamma_0-1$). Such initial profiles correspond to linear filamentation theory, where the background plasma experiences only weak density modulation. They can be easily generalized to the nonlinear case by choosing more complex profiles.

For the beam particles we take $n_{\rm b\pm}=n_{\rm b0}(1\mp\alpha_{\rm b}\cos\eta')$, $u_{{\rm b}x\pm}\approx u_{\rm b0}$ and $u_{{\rm b}y\pm}=0$, where $\alpha_{\rm b}$ is close to unity. Therefore, the initial amplitudes of the electric fields are determined approximately by
\begin{equation}
\begin{split}
&\varepsilon_{y}=2\frac{\omega_p}{kc}(n_0\alpha-n_{\rm b0}\alpha_{\rm b})\sin\eta', \\ &\varepsilon_{x}=2\frac{\omega_p}{kc}n_0\zeta \cos\eta',
\end{split}
\end{equation}
where for the initial transverse field the Gauss law is used, and for the initial longitudinal field we used equation~(\ref{elec}).

For the magnetic field, it is not possible to write such simple relations, since $j_x$ includes beam velocities. Variations in the background plasma velocities proceed in a highly asymmetric manner, reflecting the extreme sensitivity of the Lorentz factor (i.e., particle inertia) to relativistic motion at $v_{\pm x}\simeq c$. The $x$-components of velocities can be found as
\begin{equation}
    u_{\pm x}=-\left(1-u_{\pm y}^2-\frac{1}{\Gamma_\pm^2}\right)^{1/2}\!\!\!. 
\end{equation}
Therefore, for the initial magnetic field one has to numerically solve the equation
\begin{equation}\label{m}
    \frac{\partial b_z}{\partial \eta}=n_+u_{+x}-n_-u_{-x}+(n_{\rm b+}-n_{\rm b-})u_{\rm b0}+\frac{\gamma_w}{\omega_p}\varepsilon_x,
\end{equation}
with the boundary condition $b_z=0$ at $\eta'=0$.

A fourth order Runge-Kutta method was used to solve the system of equations. To use periodic boundary conditions for all quantities, we considered a segment $[-\pi/2,3\pi/2]$ for $\eta'$ that accommodates two filaments of opposite charge: the positron-dominated filament centered at $\eta'=0$ and the electron-dominated filament centered at $\eta'=\pi$. Periodic conditions are taken into account in the finite difference operator at the edges of the computational domain. For the calculation, we used finite differences of 6th order accuracy, $\sim\mathcal{O}[(\Delta\eta)^6]$, to approximate the spatial derivative with respect to $\eta$. The timestep and the cell size are $\Delta\tau=10^{-3}$, $\Delta\eta'=7\times 10^{-3}$.

\begin{figure*}
\centering
\begin{minipage}[t]{0.47\linewidth}
    \centering
    \includegraphics[width=\linewidth]{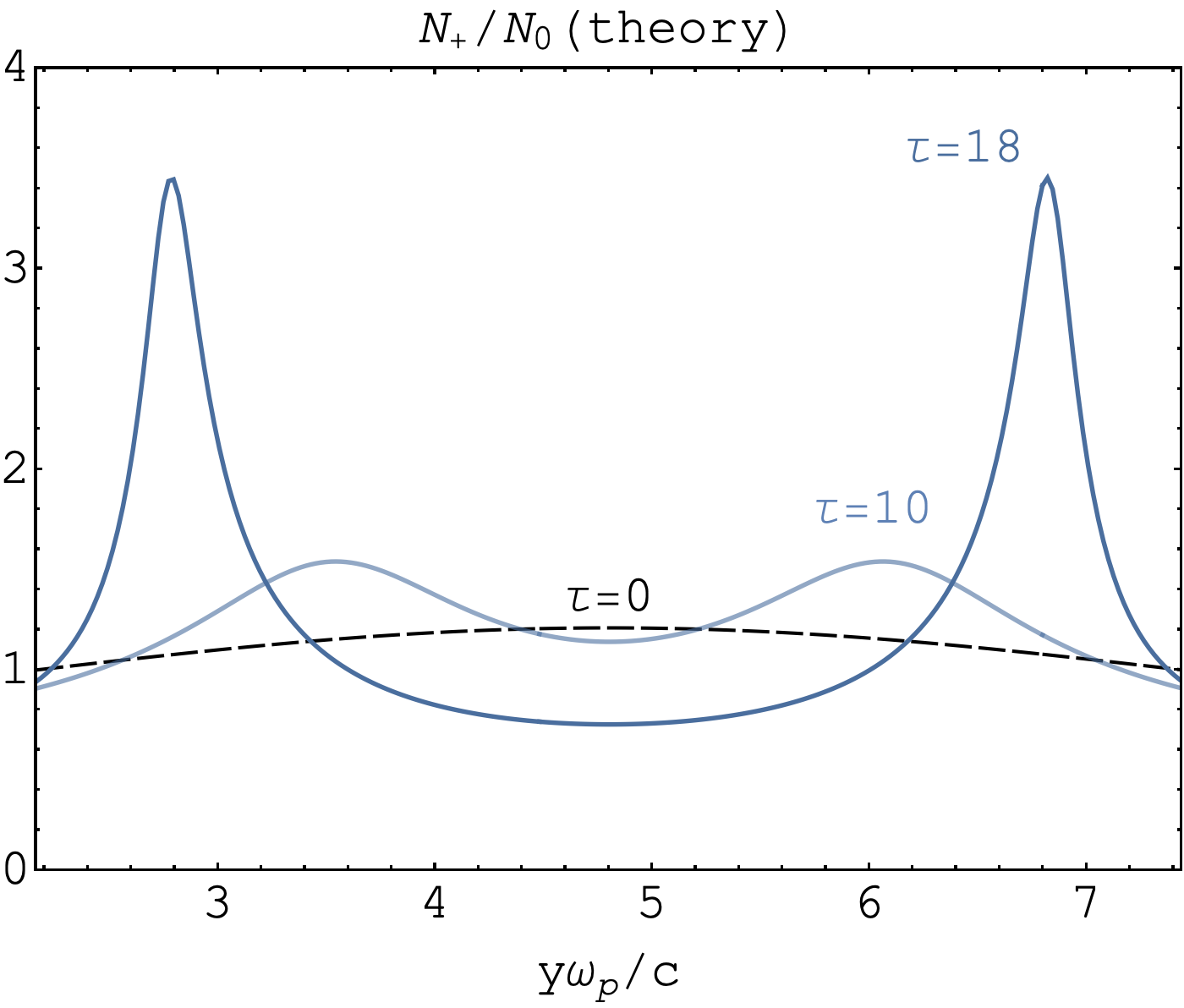}
\end{minipage}\hfill
\begin{minipage}[t]{0.5\linewidth}
    \centering
    \includegraphics[width=\linewidth]{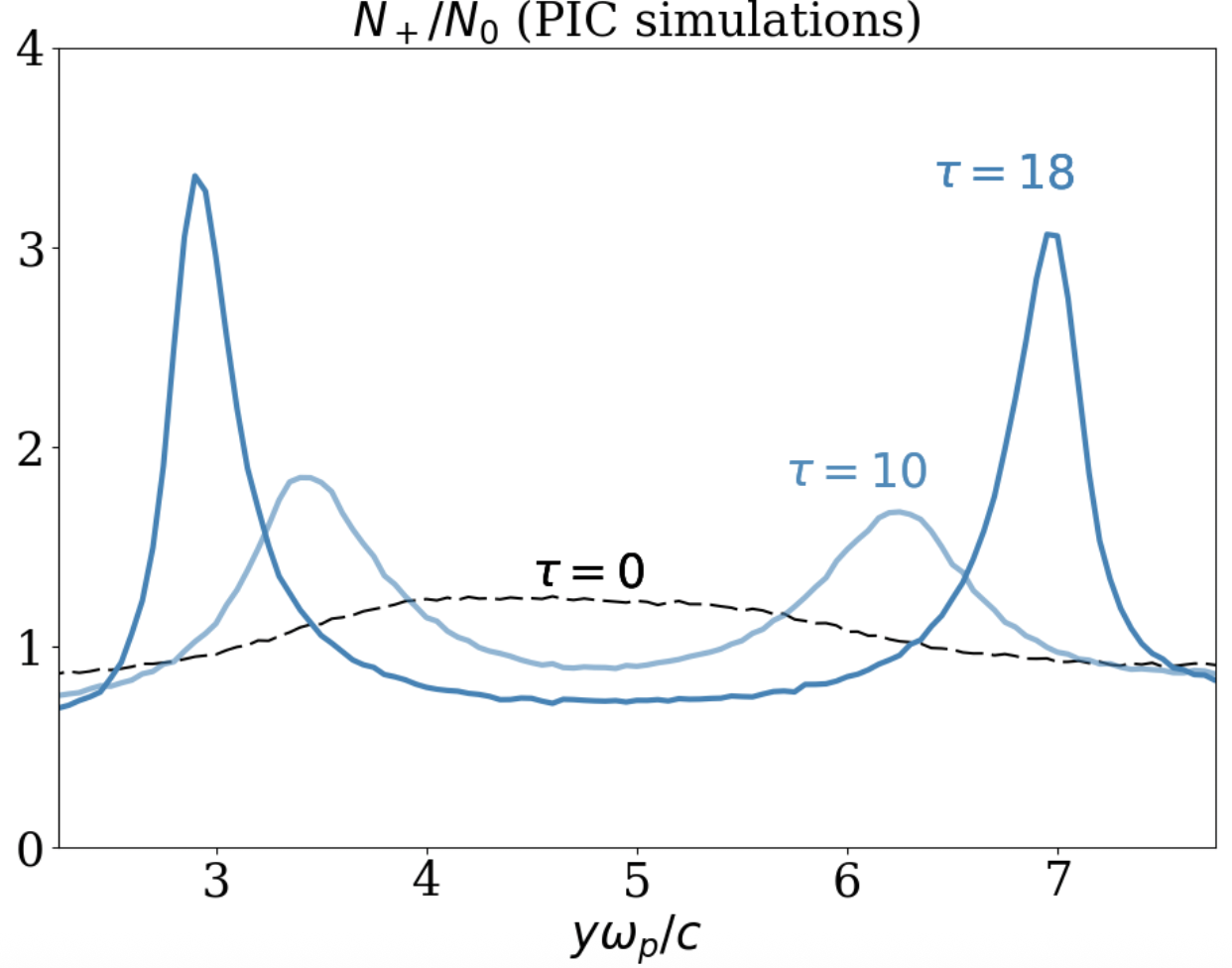}
\end{minipage}
\caption{Evolution of the positron density $n_+(\eta)$ in the lab frame $S$: left — toy model; right — reference PIC run (dimensionless time measured from $t=90\,\omega_p^{-1}$).}
\label{fig:nlab-combined}
\end{figure*}

The use of the Euler equation itself does not mean that the plasma is magnetized. It is simply a cold fluid approximation. Magnetization is determined by the time/length ratios, not by whether the model is fluid. We use the local magnetization criterion $r_{L|w}<R$, where $r_{L|w}$ is the Larmor radius of a particle in the Weibel frame, in which the electrostatic field vanishes. In Appendix~\ref{app:magnetization}, we have shown that this criterion is reduced to the inequality
\begin{equation}\label{whenM}
    b_{\rm max}>\frac{4}{\pi}\left(\frac{ck}{\omega_p}\right)\Gamma_0\Gamma_w^2\left(\sqrt{1-\frac{1}{\Gamma_0^2}}-\sqrt{1-\frac{1}{\Gamma_w^2}}\right).
\end{equation}
where $b_{\rm max}$ is the magnetic field amplitude.
This dependence on $\Gamma_w$ arises because the Weibel and upstream frames do not coincide ($\Gamma_w\neq\Gamma_0$); in the Weibel frame, the cold plasma has a finite drift, hence a nonzero Larmor radius $r_{L|w}$.

If the inequality~(\ref{whenM}) still fails before cavities are formed, particles are deflected as they stream through the filament, and are injected straight into the magnetic field extrema, where they eventually become magnetized. If the inequality is satisfied before cavity formation, plasma particles pile up and compress $B_z$ but cannot reach its extrema; the frozen-in condition blocks penetration into the strong-field region (see Fig.~\ref{fig:lab2}). Nevertheless, compression heats the plasma; as the Larmor radius grows, particles can again reach the maximum of the magnetic field $B_z$.

\subsection{Comparison with PIC simulations}\label{case1} 

In this subsection, we consider these two cases: when the plasma is magnetized before the cavity is formed and when it is not (see inequality~\ref{whenM}). In the second case, the model lies outside its validity range, but it yields qualitatively correct results.

\subsubsection{Case 1} 
The first case, motivated by shock simulations is addressed by our reference run, which is described in Section~\ref{sec:2}. The initial, $t=0$, dimensionless densities in the lab frame $S$ are chosen as $n_0=\Gamma_0/2=5$ for the plasma and $n_{\rm b0}=0.03$ for the beam\footnote{The value $n_0=\Gamma_0/2$ is related to the fact that in the SMILEI code the density normalization is taken as $N_+=m\omega_p^2n/(4\pi e^2)$, so $n=n_0$ gives $N_+(0)=N_0$.}. Therefore, the density ratio is $n_{\rm b0}/n_0=6\times10^{-3}$. The corresponding Lorentz factors are $\Gamma_0=10$ and $\Gamma_{\rm b}=3$.
The initial plasma proper temperature is $T_{\rm p}=10^{-3}mc^2$, the beam temperature is $T_{\rm b}=10mc^2$.

The initial filament radius in the PIC simulation is $R\approx 1.6(c/\omega_p)$, from which we determine the dominant wave number $k=\pi/2R\approx(\omega_p/c)$ and $\omega_p/kc\approx1$. The growth rate of the instability can be measured in simulations as
\begin{equation}
\frac{\gamma_w}{\omega_p}=\frac{1}{2}\frac{\text{d}}{\text{d}\tau}\ln\left[\frac{\langle\delta b^2_z(\tau)\rangle}{\langle\delta b^2_z(0)\rangle}\right],
\end{equation}
giving $\gamma_w\simeq0.088\omega_p$ during the linear stage. We only need the growth rate value to determine the initial magnetic field $b_z$ just before the cavity is formed (see equation~\ref{m}). Saturation happens at $t\sim 90\omega_p^{-1}$; in our model, this time corresponds to the beginning of cavity formation, $\tau=0$.

\begin{figure*}
\centering
\begin{minipage}[t]{0.49\linewidth}
    \centering
    \includegraphics[width=\linewidth]{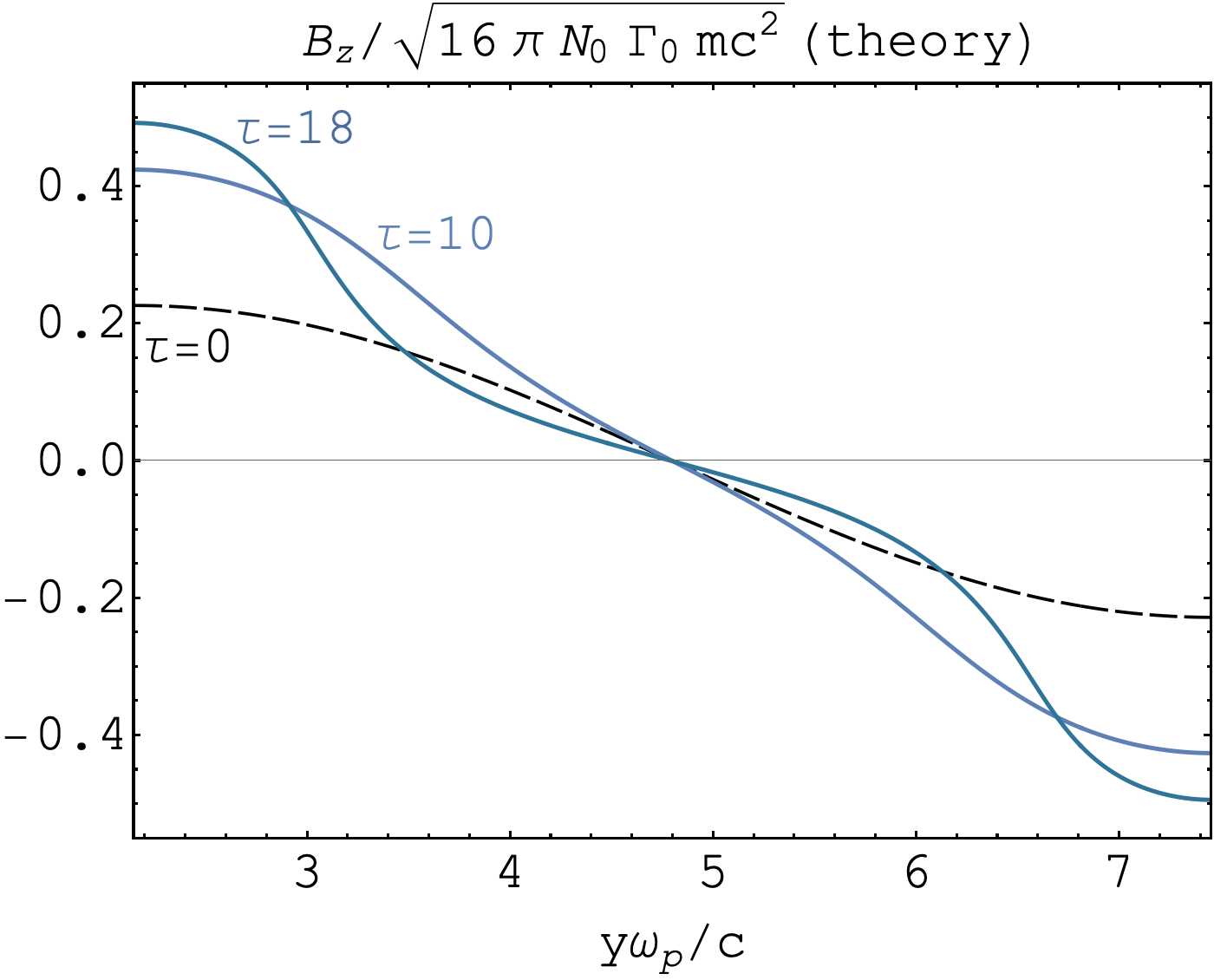}
\end{minipage}\hfill
\begin{minipage}[t]{0.5\linewidth}
    \centering
    \includegraphics[width=\linewidth]{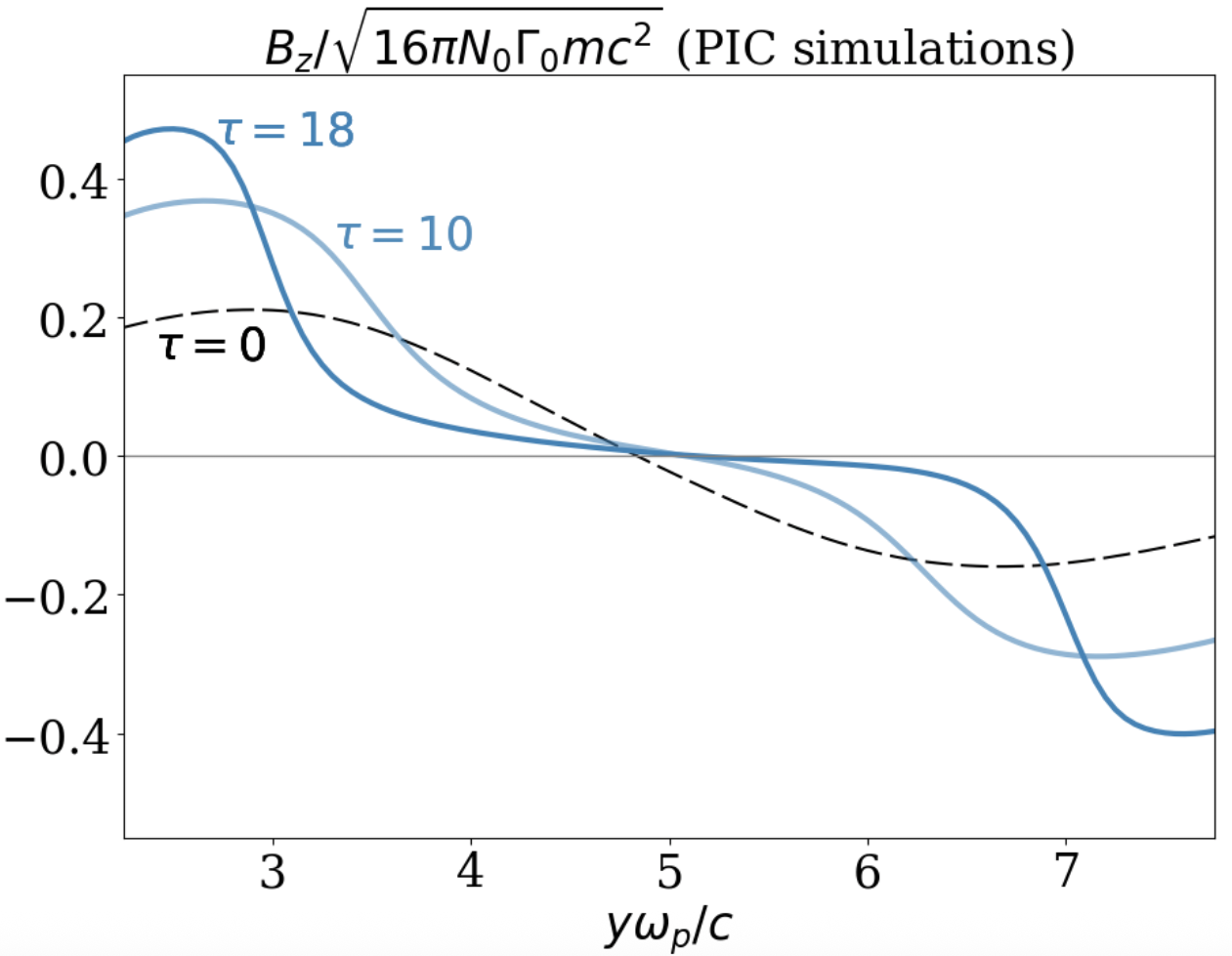}
\end{minipage}
\caption{Same as Figure~\ref{fig:nlab-combined} but for $b_z(\eta)$.}
\label{fig:Bzlab-combined}
\end{figure*}

\begin{figure*}
\centering
\begin{minipage}[t]{0.47\linewidth}
    \centering
    \includegraphics[width=\linewidth]{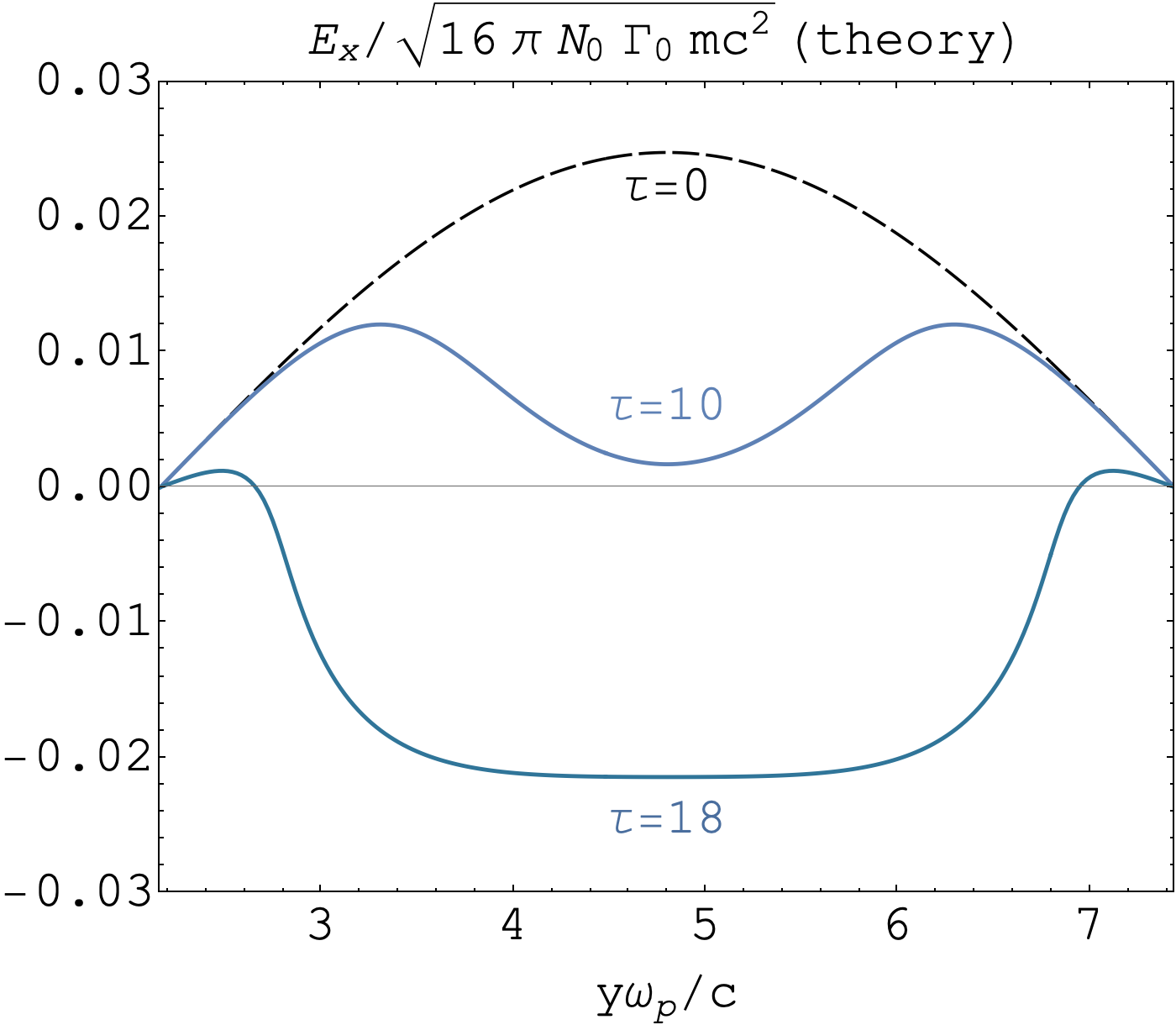}
\end{minipage}\hfill
\begin{minipage}[t]{0.5\linewidth}
    \centering
    \includegraphics[width=\linewidth]{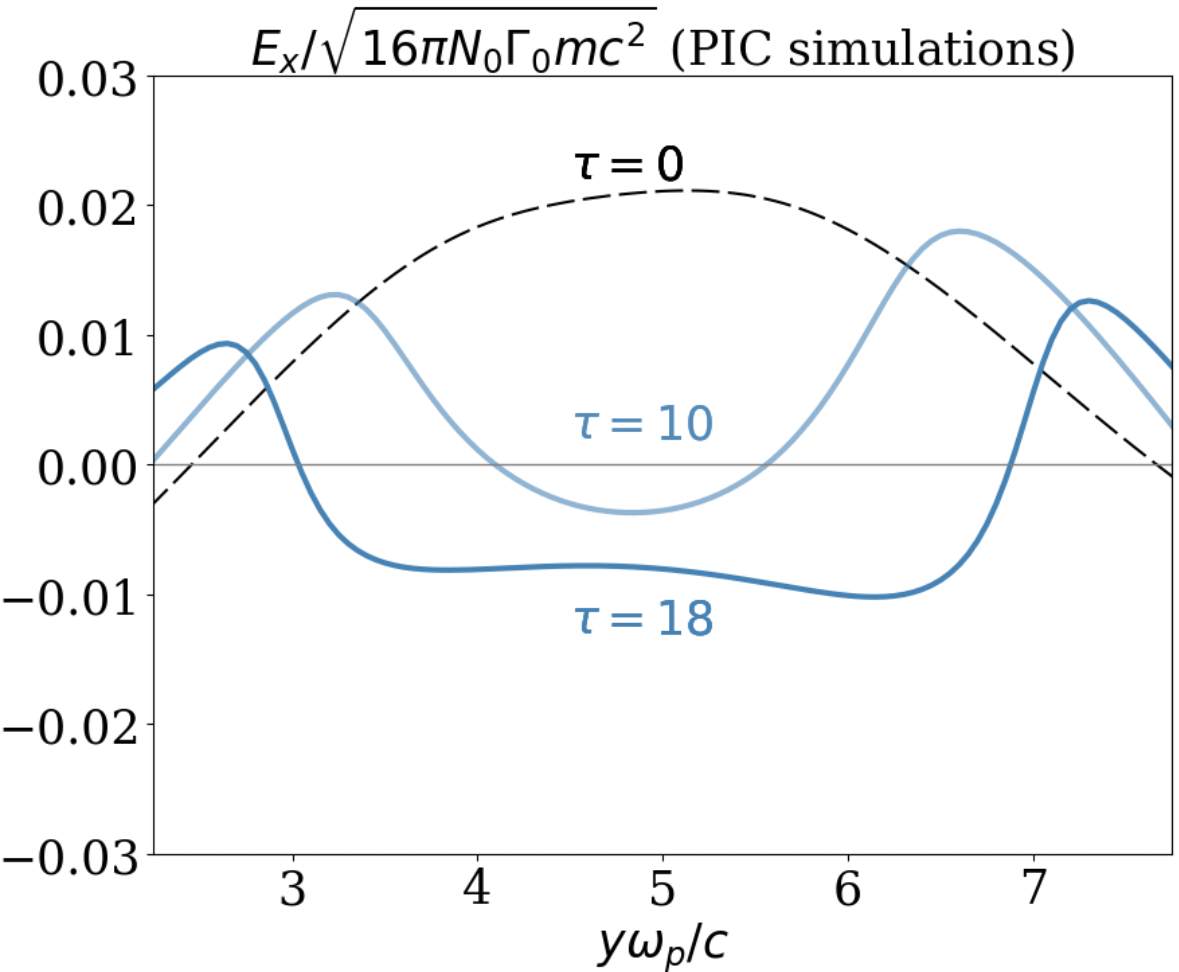}
\end{minipage}
\caption{Same as Figure~\ref{fig:nlab-combined} but for $\varepsilon_x(\eta)$.}
\label{fig:Ex-combined}
\end{figure*}

During cavity formation, the filament expands from the initial radius $R\simeq1.6(c/\omega_p)$ to $R\simeq 2.6(c/\omega_p)$. Our mathematical model does not include filament expansion; thus, for simplicity, we use the maximal radius in the calculations, which yields $\omega_p/kc=1.65$. In the calculations we adopt the following initial parameters, taken from the reference PIC run: the modulation of the plasma density is $\alpha=0.25$, the modulation of the plasma Lorentz factor is $\xi=0.3$, the modulation of the $y$-component of the plasma velocity is $\zeta=0.025$, and the analogous quantities for the beam are $\alpha_{\rm b}=0.9$, $\xi_{\rm b}=0$ and $\zeta_{\rm b}=0$.
For convenience of comparison with PIC simulations, we have translated the dimensionless transverse coordinate $\eta'=(kc/\omega_p)\eta$ into $y\omega_p/c$ by shifting the origin to $y\omega_p/c\simeq4.8$. When plotting quantities from the simulation data, we averaged them along the $x$-axis.

In all the figures, we show only the positron filament, contrasting PIC results with the simple model in the frame $S$. Comparing the two panels in Fig.~\ref{fig:nlab-combined}, we see that the maximum positron densities in our model and in PIC simulations are comparable. However, in the simulations, the filaments expand, which is not the case in our purely symmetric mathematical model. This behavior arises because different filaments are not strictly identical: small differences in their evolution cause some of them to expand, while others, formed with slight time shifts, undergo compression.

One of the remarkable properties of the cavities is not only dense walls, but also strong suppression of magnetic and transverse electric fields in the middle of the cavity. The evolution of the magnetic field profile is shown in Fig.~\ref{fig:Bzlab-combined}. The drop in the central magnetic field is stronger in the simulations, which is again associated with the expansion of the filament. Since $|E_y|\sim |B_z|$, the transverse electric field evolution is practically the same.

The longitudinal electric field evolution is shown in Fig.~\ref{fig:Ex-combined}. One can compare the behavior of this field with the behavior of the positron density (see Fig.~\ref{fig:nlab-combined}). According to equation~(\ref{rhoE}), the second derivative $\partial^2 E_x/\partial\eta^2$ is connected with the increase or decrease of charge density in the filament. In the considered filament, the plasma electron density is slightly diminished, and the beam density is negligibly small; therefore, the total charge density is regulated mostly by positrons. One can see that the positron charge density maxima correspond to the region $\partial^2E_x/\partial\eta^2<0$; this region shifts in time towards to the edges of the filament.

In this run $\Gamma_w\simeq 7$ when the condition $\Gamma_+=\Gamma_w$ is reached (see Fig.~\ref{gammaW}). The magnetization criterion~(\ref{whenM}) then becomes $b_{\rm max}\gtrsim2.6$. Since $B_z/\sqrt{16\pi N_0\Gamma_0mc^2}=b_z/(\sqrt{2}\Gamma_0)$ from Fig.~\ref{fig:Bzlab-combined} we see that the initial dimensionless amplitude of magnetic field is $b_{\rm max}\simeq 3.2$. Therefore, the plasma is magnetized before it accumulates onto the walls, and the plasma motion compresses the magnetic field. The dense plasma walls of adjacent filaments contain a strong magnetic field between them, creating a kind of "magnetic sandwich". This pattern is also observed with other parameters of the beams (see Table~\ref{tab1}).

\subsubsection{Case 2} 
\begin{figure}
    \begin{minipage}{0.48\textwidth}
        \centering
        \includegraphics[width=\linewidth]{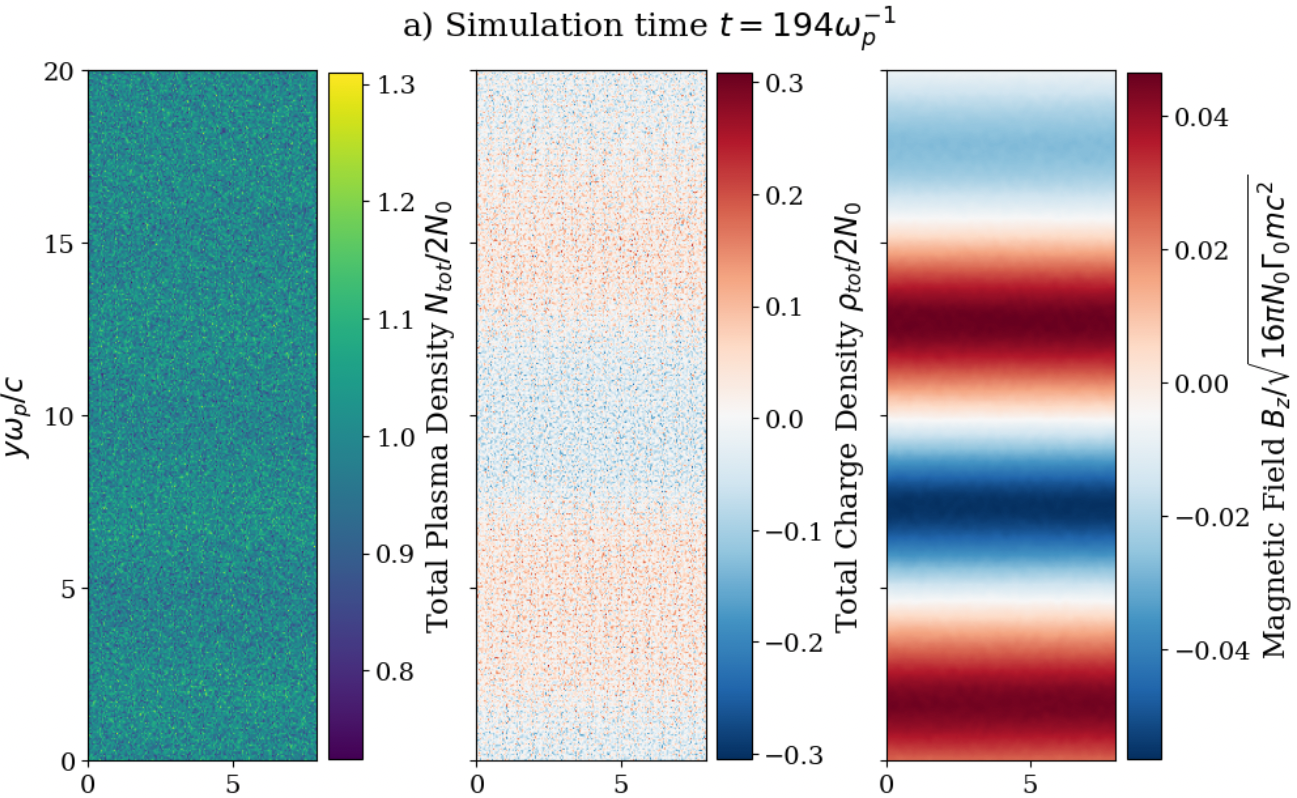}
        \label{fig:case1}
    \end{minipage}%
    \hfill
    \begin{minipage}{0.48\textwidth}
        \centering
        \includegraphics[width=\linewidth]{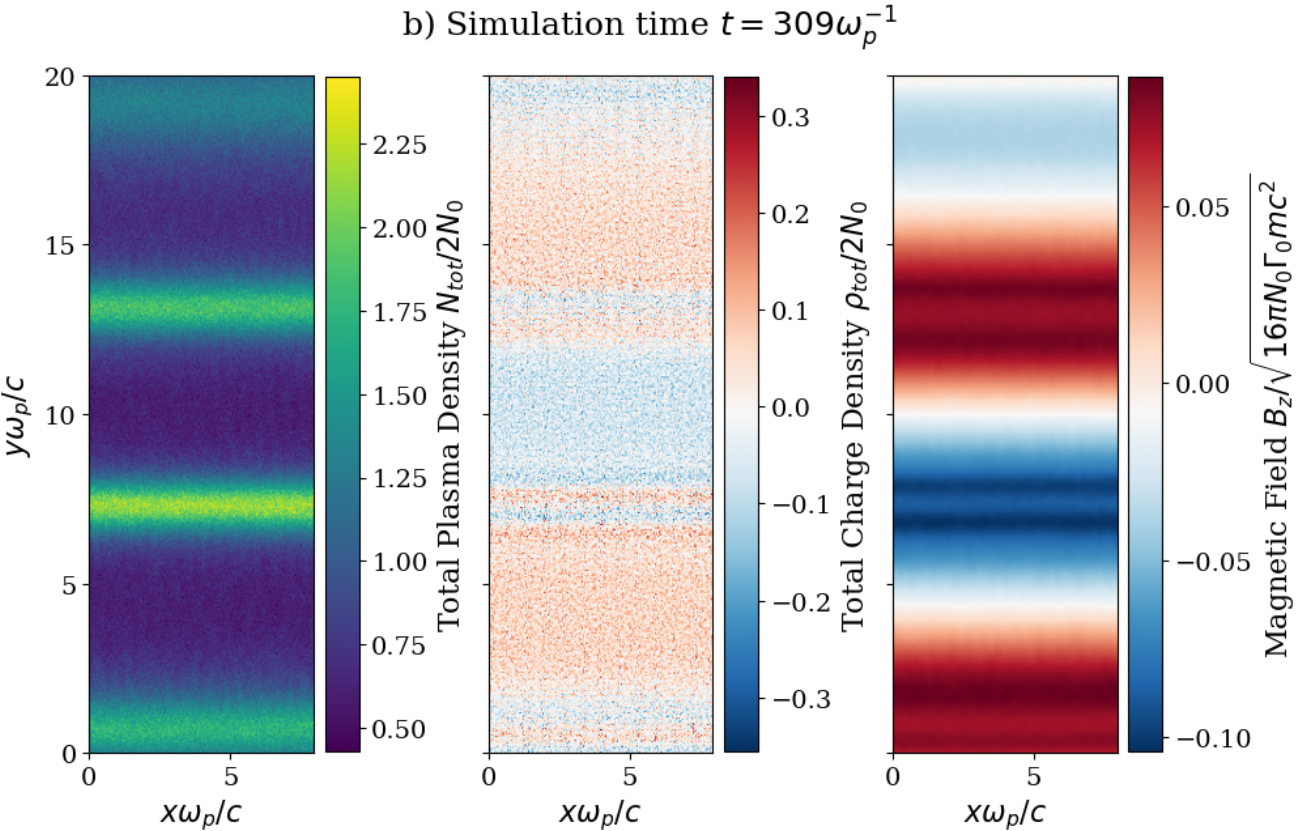}
        \caption{
        Simulation setup 2.
        Distributions of total plasma density, total charge density, and magnetic field $B_z$ in the lab frame $S$ at different simulation times: a) just before cavity formation; b) when cavities are formed. The figure shows only the part of the simulation box, $y<20(c/\omega_p)$, where the cavity walls are the strongest.}
        \label{fig:case2}
    \end{minipage}
    \end{figure}

Let the initial densities in the $S$ frame be $n_0=5$ and $n_{\rm \rm b0}=0.01$, such that $n_{\rm \rm b0}/n_{0}=2\times10^{-3}$. All other parameters are kept the same as in the reference run. This setup corresponds to run 2 in Table~\ref{tab1}. According to PIC simulations, cavity formation ($\Gamma_+=\Gamma_w\simeq 9$ at the filament center) starts at $\sim 190\omega_p^{-1}$ and lasts approximately $\sim 100\omega_p^{-1}$ (see Fig.~\ref{fig:case2}). The initial radius of the filament is $R\simeq 3(c/\omega_p)$ and remains nearly constant during cavity formation. It gives $\omega_p/ck\simeq 1.9$.

The growth rate in this PIC simulation is $\gamma_w\simeq 0.032\omega_p$. According to simulation results, the initial parameters are $\alpha=0.03$, $\xi=0.1$, $\zeta=0.0025$, and $\alpha_{\rm b}=0.4$. Thus, unlike the reference simulation, where $\alpha_{\rm b}\simeq 0.9$, here the formation of a cavity begins even before the beam is completely saturated (second scenario of cavity formation, see Section~\ref{qual}). 

\begin{figure*}
\centering
\begin{minipage}[t]{0.47\linewidth}
    \centering
    \includegraphics[width=\linewidth]{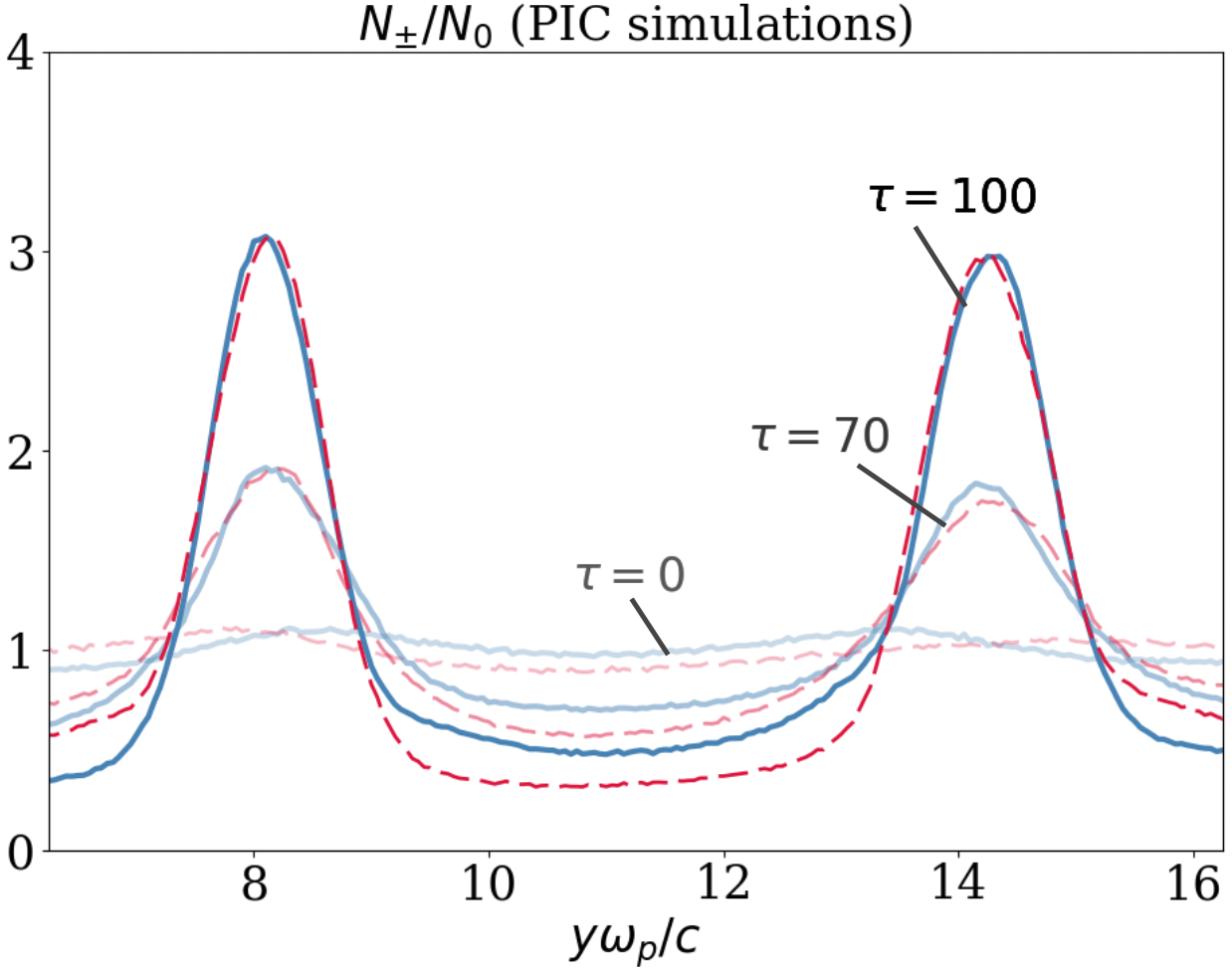}
\end{minipage}\hfill
\begin{minipage}[t]{0.5\linewidth}
    \centering
    \includegraphics[width=\linewidth]{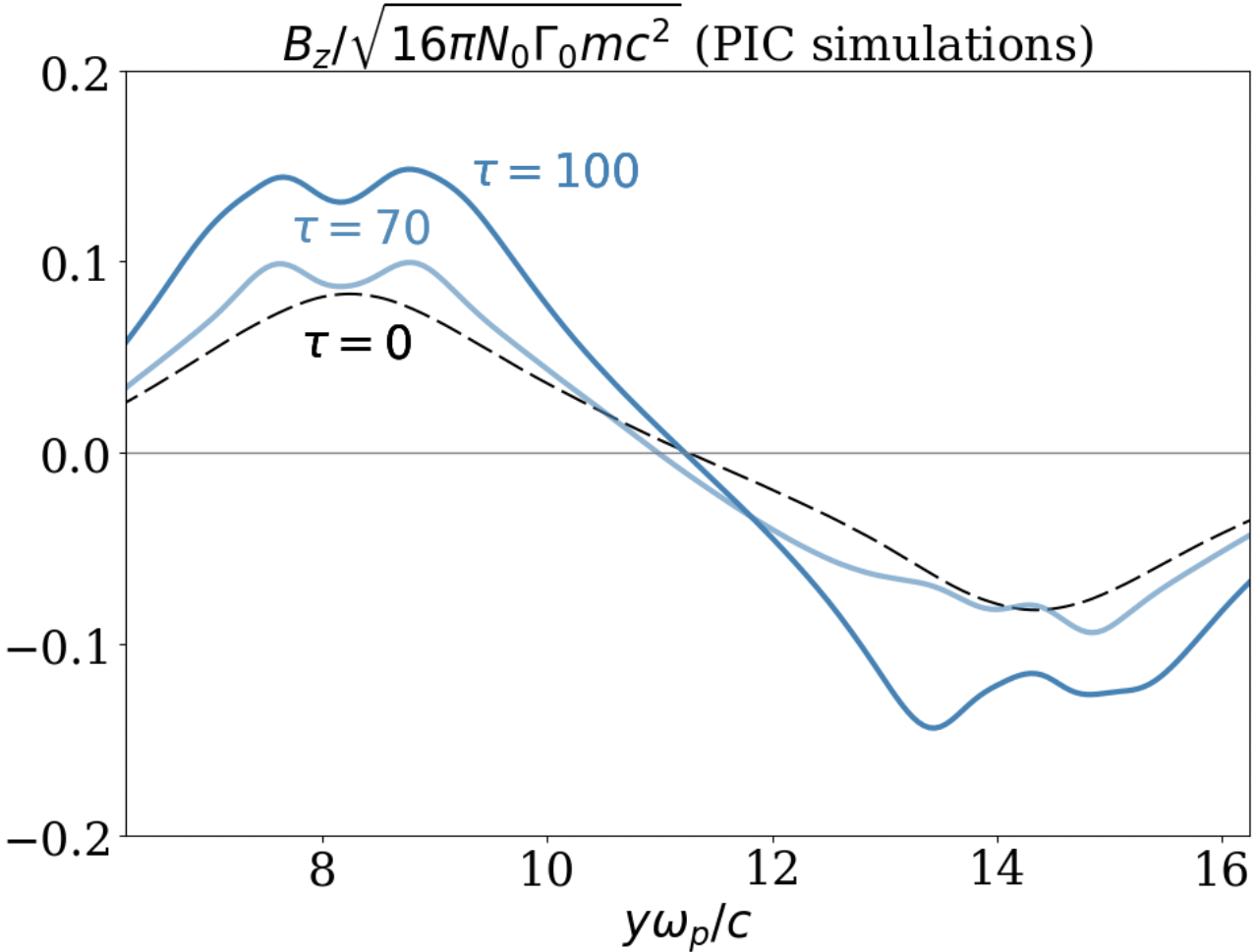}
\end{minipage}
\caption{Structure of a positron-dominated filament in simulation setup 2: left — evolution of the positron density $n_+(\eta)$ (blue, solid) and electron density $n_-(\eta)$ (red, dashed) according to PIC simulations; right — evolution of the magnetic field $b_z(\eta)$ according to PIC simulations (run 2). The dimensionless time $\tau$ is measured from $t=190\,\omega_p^{-1}$.}
\label{fig:pic1-combined}
\end{figure*}

\begin{figure*}
\centering
\begin{minipage}[t]{0.46\linewidth}
    \centering
    \includegraphics[width=\linewidth]{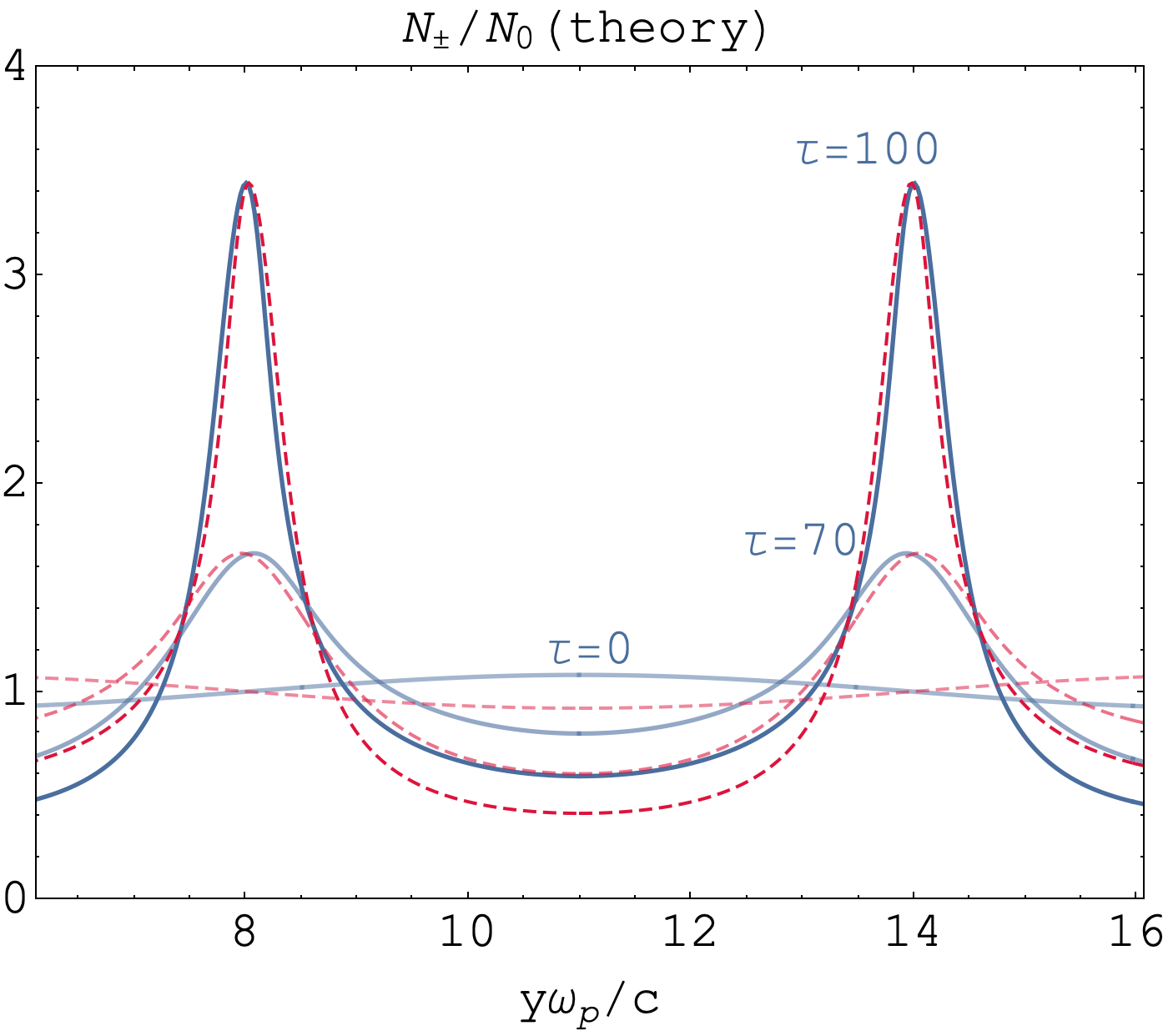}
\end{minipage}\hfill
\begin{minipage}[t]{0.52\linewidth}
    \centering
    \includegraphics[width=\linewidth]{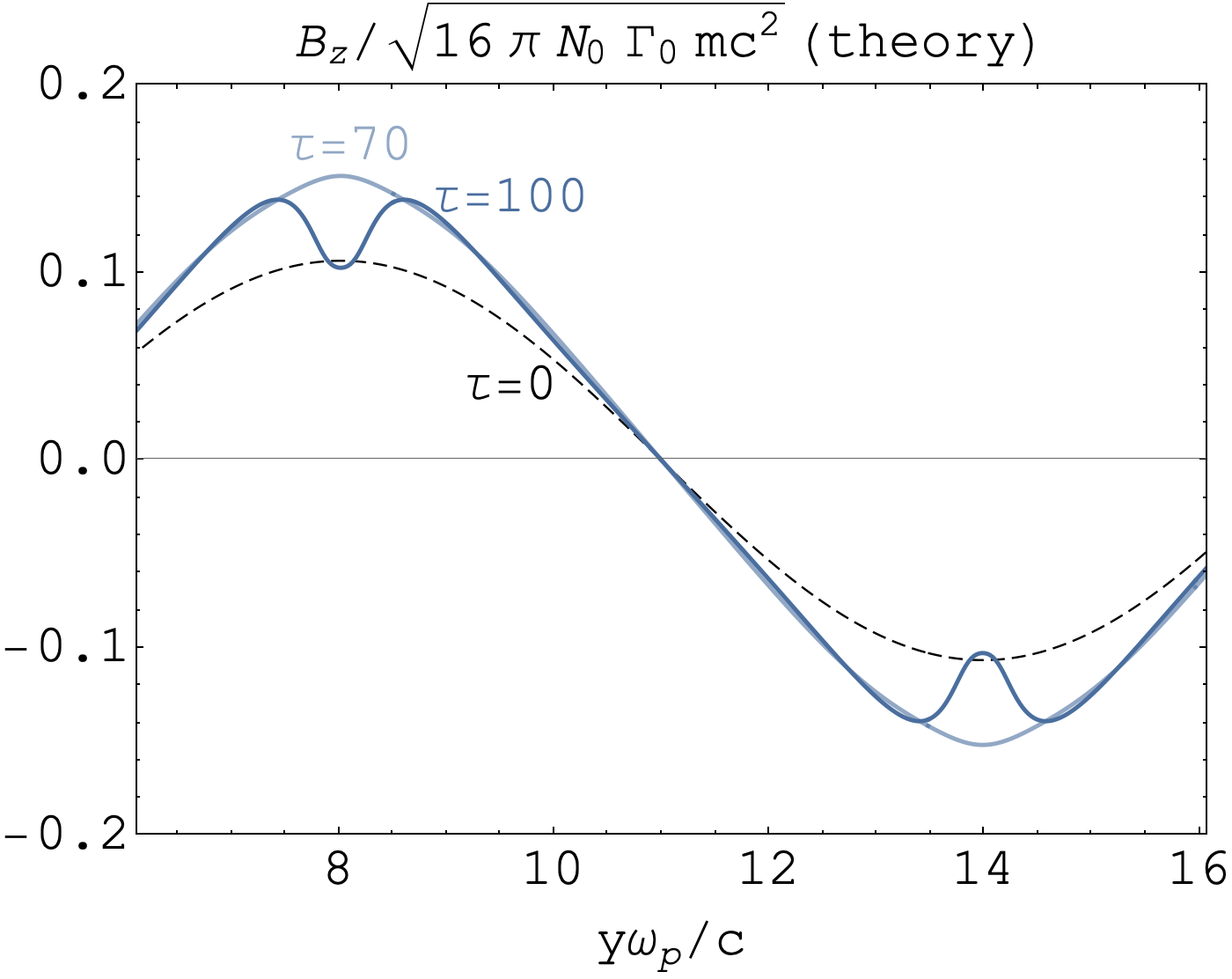}
\end{minipage}
\caption{Structure of a positron-dominated filament in simulation setup 2: left — evolution of the positron density $n_+(\eta)$ (blue, solid) and electron density $n_-(\eta)$ (red, dashed) according to the simple model; right — evolution of the magnetic field $b_z(\eta)$ according to the toy model.}
\label{fig:model1-combined}
\end{figure*}

The magnetic field amplitude at the onset of cavity formation is $b_{\rm max}\simeq 0.5$, whereas the magnetization criterion~(\ref{whenM}) requires $b_{\rm max}\gtrsim 0.64$, suggesting a non-magnetized plasma. 
The Larmor radius of the particles is large and they are simply deflected toward the edges of the filament. The plasma accumulates at these edges, which is clearly seen in the density plot in Fig.~\ref{fig:pic1-combined}: the solid blue curve (plasma positrons) and the dashed red curve (plasma electrons) both develop maxima at the initial filament boundaries.
Two $e^{\pm}$ density peaks (within a cavity wall) execute small relative transverse oscillations in anti-phase. Opposite charges within the walls nearly cancel each other, keeping $\rho$ and $\boldsymbol{j}$ small. At maximum charge separation the cancellation weakens, leaving a net charge and a wall current. Cross-mixing ($e^+$ entering the $e^-$ filament and vice versa) yields a local minimum of $B$ between cavities (see $\tau=100$ in Fig.~\ref{fig:model1-combined}).
Conversely, when positrons and electrons remain within their own filaments, a local magnetic maximum is formed (see $\tau=70$ in Fig.~\ref{fig:model1-combined}). Thus, due to the compression of the cavity walls and the oscillatory displacement of charges within them, the magnetic field extrema
also oscillates. Since these wall currents are weak, they do not contribute significantly to the overall magnetic field. Instead, the magnetic field is mainly sustained by the dilute beam and the remaining background plasma inside the cavity.

The evolved configuration presents plasma in the walls, surrounded by an enhanced magnetic field, which we refer to as a "plasma sandwich". Similar behavior is also seen in simulations of other parameters (see Table~\ref{tab1}).

Although our simple model is not suitable for describing this case, we can qualitatively obtain a similar result by artificially increasing the beam modulation to $\alpha_{\rm b}\simeq 0.7$. To be consistent, we have to increase the plasma modulation to $\alpha=0.8$ and its deceleration to $\xi=0.145$ (this gives $\Gamma_w\sim 9$ and compensation of electric and magnetic forces for positrons). The transverse velocity modulation is kept the same, $\zeta=0.025$. The model also shows that the walls grow directly at the filament boundary, inside the magnetic field maxima. However, the walls are thinner, and the magnetic field reaches saturation faster (compare $\tau=70$ and $\tau=100$ in Fig.~\ref{fig:pic1-combined} and Fig.~\ref{fig:model1-combined}).

\section{Discussion and conclusions}\label{sec:4}

In relativistic Weibel shocks, such cavities appear naturally owing to the low number density of returning particles in comparison with the upstream flow. It is generally accepted that cavities form at sufficiently late simulation times. However, they also form at early stages of the shock evolution. The distance from the shock front at which they are formed increases with time. 

The reason cavities are expected to appear only at late times is that their significant expansion is not possible in a perfectly symmetric configuration, as in the early stages of shock simulations. If there is an equal number of filaments carrying opposite equal currents, no net expansion can occur. For significant expansion to take place, this symmetry must be broken. Factors that can influence symmetry breaking include plasma turbulence in the precursor and a high temperature of returning particles. Simulations also demonstrate that in relativistic shocks, the cross-section–averaged densities of returning electrons and positrons are not equal (\citealt{Groselj2024}). In different regions of the shock precursor, one type of returning particle dominates over the other. Presumably, the dominance of one type of particle species and its inhomogeneity along the $x$-axis leads to an expansion of the cavities of the appropriate type and a decrease in the size of the cavities of the other type, similar to the mechanism described by \cite{Reville2012}. 

The main conclusions are as follows. We separate the process of formation of the cellular structure of the shock precursor (cavities separated by dense plasma walls) from the subsequent expansion of the cavities, the thickening of their walls, and the conversion of these walls to long-lasting structures downstream. 
In different inertial frames, both the mechanism of cavity formation and the observed cavity structure can look different due to the Lorentz transformation of fields. It is therefore convenient to analyze the process in a particular (special) frame, e.g., the Weibel frame, in which the electrostatic field vanishes. In this frame, a cavity forms when the inductive electric field associated with the growing magnetic field can reverse the particle motion; the magnetic force then changes sign and expels particles from the filament center.
This mechanism operates self-consistently in homogeneous beam–plasma systems, without requiring large-scale shock inhomogeneities. The inhomogeneity of the returning particles contributes to the expansion of the cavities (e.g. \citealt{Peterson2021, Peterson2022}).
We conclude that the cavitation instability is an intrinsic outcome of asymmetric beam–plasma interactions in relativistic shocks. It bridges the gap between kinetic-scale Weibel turbulence and macroscopic magnetic structures, thereby offering a compelling step towards the solution to the long-standing GRB afterglow magnetization problem.

\newpage

\section*{Acknowledgements}

We acknowledge useful discussions with Julia Kropotina, Vadim Romansky, Kuan-Chou Hou, Arka Ghosh, and Daniel Groselj. We are grateful to the SMILEI team for their prompt feedback and insightful comments. This research was supported by the Israel Science Foundation under grants 2067/19 and 2126/22.




\bibliographystyle{mnras}
\bibliography{References} 

\begin{thebibliography}{}
\makeatletter
\relax
\def\mn@urlcharsother{\let\do\@makeother \do\$\do\&\do\#\do\^\do\_\do\%\do\~}
\def\mn@doi{\begingroup\mn@urlcharsother \@ifnextchar [ {\mn@doi@} {\mn@doi@[]}}
\def\mn@doi@[#1]#2{\def\@tempa{#1}\ifx\@tempa\@empty \href {http://dx.doi.org/#2} {doi:#2}\else \href {http://dx.doi.org/#2} {#1}\fi \endgroup}
\def\mn@eprint#1#2{\mn@eprint@#1:#2::\@nil}
\def\mn@eprint@arXiv#1{\href {http://arxiv.org/abs/#1} {{\tt arXiv:#1}}}
\def\mn@eprint@dblp#1{\href {http://dblp.uni-trier.de/rec/bibtex/#1.xml} {dblp:#1}}
\def\mn@eprint@#1:#2:#3:#4\@nil{\def\@tempa {#1}\def\@tempb {#2}\def\@tempc {#3}\ifx \@tempc \@empty \let \@tempc \@tempb \let \@tempb \@tempa \fi \ifx \@tempb \@empty \def\@tempb {arXiv}\fi \@ifundefined {mn@eprint@\@tempb}{\@tempb:\@tempc}{\expandafter \expandafter \csname mn@eprint@\@tempb\endcsname \expandafter{\@tempc}}}

\bibitem[\protect\citeauthoryear{{Blinne}, {Schinkel}, {Kuschel}, {Elkina}, {Rykovanov}  \& {Zepf}}{{Blinne} et~al.}{2018}]{Blinne2018}
{Blinne} A.,  {Schinkel} D.,  {Kuschel} S.,  {Elkina} N.,  {Rykovanov} S.~G.,   {Zepf} M.,  2018, \mn@doi [Computer Physics Communications] {10.1016/j.cpc.2017.10.010}, \href {https://ui.adsabs.harvard.edu/abs/2018CoPhC.224..273B} {224, 273}

\bibitem[\protect\citeauthoryear{{Brainerd}}{{Brainerd}}{2000}]{Brainerd2000}
{Brainerd} J.~J.,  2000, in {Kippen} R.~M.,  {Mallozzi} R.~S.,   {Fishman} G.~J.,  eds,  American Institute of Physics Conference Series Vol. 526, Gamma-ray Bursts, 5th Huntsville Symposium. AIP, pp 455--459 (\mn@eprint {arXiv} {astro-ph/9904040}), \mn@doi{10.1063/1.1361580}

\bibitem[\protect\citeauthoryear{{Bresci}, {Gremillet}  \& {Lemoine}}{{Bresci} et~al.}{2022}]{Bresci2022}
{Bresci} V.,  {Gremillet} L.,   {Lemoine} M.,  2022, \mn@doi [\pre] {10.1103/PhysRevE.105.035202}, \href {https://ui.adsabs.harvard.edu/abs/2022PhRvE.105c5202B} {105, 035202}

\bibitem[\protect\citeauthoryear{{Califano}, {Cecchi}  \& {Chiuderi}}{{Califano} et~al.}{2002}]{Califano2002}
{Califano} F.,  {Cecchi} T.,   {Chiuderi} C.,  2002, \mn@doi [Physics of Plasmas] {10.1063/1.1435001}, \href {https://ui.adsabs.harvard.edu/abs/2002PhPl....9..451C} {9, 451}

\bibitem[\protect\citeauthoryear{{Chang}, {Spitkovsky}  \& {Arons}}{{Chang} et~al.}{2008}]{Chang2008}
{Chang} P.,  {Spitkovsky} A.,   {Arons} J.,  2008, \mn@doi [\apj] {10.1086/524764}, \href {https://ui.adsabs.harvard.edu/abs/2008ApJ...674..378C} {674, 378}

\bibitem[\protect\citeauthoryear{{Derouillat} et~al.,}{{Derouillat} et~al.}{2018}]{Derouillat2018}
{Derouillat} J.,  et~al., 2018, \mn@doi [Computer Physics Communications] {10.1016/j.cpc.2017.09.024}, \href {https://ui.adsabs.harvard.edu/abs/2018CoPhC.222..351D} {222, 351}

\bibitem[\protect\citeauthoryear{{Fried}}{{Fried}}{1959}]{Fried1959}
{Fried} B.~D.,  1959, \mn@doi [Physics of Fluids] {10.1063/1.1705933}, \href {https://ui.adsabs.harvard.edu/abs/1959PhFl....2..337F} {2, 337}

\bibitem[\protect\citeauthoryear{{Gro{\v{s}}elj}, {Sironi}  \& {Beloborodov}}{{Gro{\v{s}}elj} et~al.}{2022}]{Groselj2022}
{Gro{\v{s}}elj} D.,  {Sironi} L.,   {Beloborodov} A.~M.,  2022, \mn@doi [\apj] {10.3847/1538-4357/ac713e}, \href {https://ui.adsabs.harvard.edu/abs/2022ApJ...933...74G} {933, 74}

\bibitem[\protect\citeauthoryear{{Gro{\v{s}}elj}, {Sironi}  \& {Spitkovsky}}{{Gro{\v{s}}elj} et~al.}{2024}]{Groselj2024}
{Gro{\v{s}}elj} D.,  {Sironi} L.,   {Spitkovsky} A.,  2024, \mn@doi [\apjl] {10.3847/2041-8213/ad2c8c}, \href {https://ui.adsabs.harvard.edu/abs/2024ApJ...963L..44G} {963, L44}

\bibitem[\protect\citeauthoryear{{Gruzinov}}{{Gruzinov}}{2001}]{Gruzinov2001}
{Gruzinov} A.,  2001, \mn@doi [\apjl] {10.1086/324223}, \href {https://ui.adsabs.harvard.edu/abs/2001ApJ...563L..15G} {563, L15}

\bibitem[\protect\citeauthoryear{{Gruzinov} \& {Waxman}}{{Gruzinov} \& {Waxman}}{1999}]{GruzinovWaxman1999}
{Gruzinov} A.,  {Waxman} E.,  1999, \mn@doi [\apj] {10.1086/306720}, \href {https://ui.adsabs.harvard.edu/abs/1999ApJ...511..852G} {511, 852}

\bibitem[\protect\citeauthoryear{{Higuera} \& {Cary}}{{Higuera} \& {Cary}}{2017}]{HigueraCary2017}
{Higuera} A.~V.,  {Cary} J.~R.,  2017, \mn@doi [Physics of Plasmas] {10.1063/1.4979989}, \href {https://ui.adsabs.harvard.edu/abs/2017PhPl...24e2104H} {24, 052104}

\bibitem[\protect\citeauthoryear{{Katz}, {Keshet}  \& {Waxman}}{{Katz} et~al.}{2007}]{Katz2007}
{Katz} B.,  {Keshet} U.,   {Waxman} E.,  2007, \mn@doi [\apj] {10.1086/509115}, \href {https://ui.adsabs.harvard.edu/abs/2007ApJ...655..375K} {655, 375}

\bibitem[\protect\citeauthoryear{{Keshet}, {Katz}, {Spitkovsky}  \& {Waxman}}{{Keshet} et~al.}{2009}]{Keshet2009}
{Keshet} U.,  {Katz} B.,  {Spitkovsky} A.,   {Waxman} E.,  2009, \mn@doi [\apjl] {10.1088/0004-637X/693/2/L127}, \href {https://ui.adsabs.harvard.edu/abs/2009ApJ...693L.127K} {693, L127}

\bibitem[\protect\citeauthoryear{{Lemoine}}{{Lemoine}}{2015}]{Lemoine2015}
{Lemoine} M.,  2015, \mn@doi [Journal of Plasma Physics] {10.1017/S0022377814000920}, \href {https://ui.adsabs.harvard.edu/abs/2015JPlPh..81a4501L} {81, 455810101}

\bibitem[\protect\citeauthoryear{{Lemoine}, {Gremillet}, {Pelletier}  \& {Vanthieghem}}{{Lemoine} et~al.}{2019}]{Lemoine2019}
{Lemoine} M.,  {Gremillet} L.,  {Pelletier} G.,   {Vanthieghem} A.,  2019, \mn@doi [\prl] {10.1103/PhysRevLett.123.035101}, \href {https://ui.adsabs.harvard.edu/abs/2019PhRvL.123c5101L} {123, 035101}

\bibitem[\protect\citeauthoryear{{Lu}, {Kilian}, {Guo}, {Li}  \& {Liang}}{{Lu} et~al.}{2020}]{Lu2020}
{Lu} Y.,  {Kilian} P.,  {Guo} F.,  {Li} H.,   {Liang} E.,  2020, \mn@doi [Journal of Computational Physics] {10.1016/j.jcp.2020.109388}, \href {https://ui.adsabs.harvard.edu/abs/2020JCoPh.41309388L} {413, 109388}

\bibitem[\protect\citeauthoryear{{Medvedev} \& {Loeb}}{{Medvedev} \& {Loeb}}{1999}]{MedvedevLoeb1999}
{Medvedev} M.~V.,  {Loeb} A.,  1999, \mn@doi [\apj] {10.1086/308038}, \href {https://ui.adsabs.harvard.edu/abs/1999ApJ...526..697M} {526, 697}

\bibitem[\protect\citeauthoryear{{Naseri}, {Bochkarev}, {Ruan}, {Bychenkov}, {Khudik}  \& {Shvets}}{{Naseri} et~al.}{2018}]{Naseri2018}
{Naseri} N.,  {Bochkarev} S.~G.,  {Ruan} P.,  {Bychenkov} V.~Y.,  {Khudik} V.,   {Shvets} G.,  2018, \mn@doi [Physics of Plasmas] {10.1063/1.5008278}, \href {https://ui.adsabs.harvard.edu/abs/2018PhPl...25a2118N} {25, 012118}

\bibitem[\protect\citeauthoryear{{Pelletier}, {Gremillet}, {Vanthieghem}  \& {Lemoine}}{{Pelletier} et~al.}{2019}]{Pelletier2019}
{Pelletier} G.,  {Gremillet} L.,  {Vanthieghem} A.,   {Lemoine} M.,  2019, \mn@doi [\pre] {10.1103/PhysRevE.100.013205}, \href {https://ui.adsabs.harvard.edu/abs/2019PhRvE.100a3205P} {100, 013205}

\bibitem[\protect\citeauthoryear{{Peterson}, {Glenzer}  \& {Fiuza}}{{Peterson} et~al.}{2021}]{Peterson2021}
{Peterson} J.~R.,  {Glenzer} S.,   {Fiuza} F.,  2021, \mn@doi [\prl] {10.1103/PhysRevLett.126.215101}, \href {https://ui.adsabs.harvard.edu/abs/2021PhRvL.126u5101P} {126, 215101}

\bibitem[\protect\citeauthoryear{{Peterson}, {Glenzer}  \& {Fiuza}}{{Peterson} et~al.}{2022}]{Peterson2022}
{Peterson} J.~R.,  {Glenzer} S.,   {Fiuza} F.,  2022, \mn@doi [\apjl] {10.3847/2041-8213/ac44a2}, \href {https://ui.adsabs.harvard.edu/abs/2022ApJ...924L..12P} {924, L12}

\bibitem[\protect\citeauthoryear{{Reville} \& {Bell}}{{Reville} \& {Bell}}{2012}]{Reville2012}
{Reville} B.,  {Bell} A.~R.,  2012, \mn@doi [\mnras] {10.1111/j.1365-2966.2011.19892.x}, \href {https://ui.adsabs.harvard.edu/abs/2012MNRAS.419.2433R} {419, 2433}

\bibitem[\protect\citeauthoryear{{Reville} \& {Bell}}{{Reville} \& {Bell}}{2014}]{RevilleBell2014}
{Reville} B.,  {Bell} A.~R.,  2014, \mn@doi [\mnras] {10.1093/mnras/stu088}, \href {https://ui.adsabs.harvard.edu/abs/2014MNRAS.439.2050R} {439, 2050}

\bibitem[\protect\citeauthoryear{{Sari}, {Piran}  \& {Narayan}}{{Sari} et~al.}{1998}]{Sari1998}
{Sari} R.,  {Piran} T.,   {Narayan} R.,  1998, \mn@doi [\apjl] {10.1086/311269}, \href {https://ui.adsabs.harvard.edu/abs/1998ApJ...497L..17S} {497, L17}

\bibitem[\protect\citeauthoryear{{Sironi}, {Spitkovsky}  \& {Arons}}{{Sironi} et~al.}{2013}]{Sironi2013}
{Sironi} L.,  {Spitkovsky} A.,   {Arons} J.,  2013, \mn@doi [\apj] {10.1088/0004-637X/771/1/54}, \href {https://ui.adsabs.harvard.edu/abs/2013ApJ...771...54S} {771, 54}

\bibitem[\protect\citeauthoryear{{Spitkovsky}}{{Spitkovsky}}{2008}]{Spitkovsky2008}
{Spitkovsky} A.,  2008, \mn@doi [\apjl] {10.1086/527374}, \href {https://ui.adsabs.harvard.edu/abs/2008ApJ...673L..39S} {673, L39}

\bibitem[\protect\citeauthoryear{{Waxman}}{{Waxman}}{1997}]{Waxman1997}
{Waxman} E.,  1997, \mn@doi [\apjl] {10.1086/310809}, \href {https://ui.adsabs.harvard.edu/abs/1997ApJ...485L...5W} {485, L5}

\bibitem[\protect\citeauthoryear{{Weibel}}{{Weibel}}{1959}]{Weibel1959}
{Weibel} E.~S.,  1959, \mn@doi [\prl] {10.1103/PhysRevLett.2.83}, \href {https://ui.adsabs.harvard.edu/abs/1959PhRvL...2...83W} {2, 83}

\makeatother
\end{thebibliography}




\appendix

\section{Boosted reference frame}\label{sec:boosted}

\subsection{Lorentz boost}

In a new reference frame, both the process of cavity formation and its internal structure appear different. Analyzing cavity formation in various reference frames provides deeper insight into this phenomenon.

Let us boost to a reference frame $S'$, in which the background plasma is at rest at $t=0$. For convenience, henceforth we designate variables in this $S'$ reference frame without a prime, i.e., as $(ct, x)$. Instead, all variables from the lab frame $S$ are labeled with an index "lab", $(c
t_{\rm lab},x_{\rm lab})$. 

\begin{figure}
\centering
\includegraphics[width=0.4\textwidth]{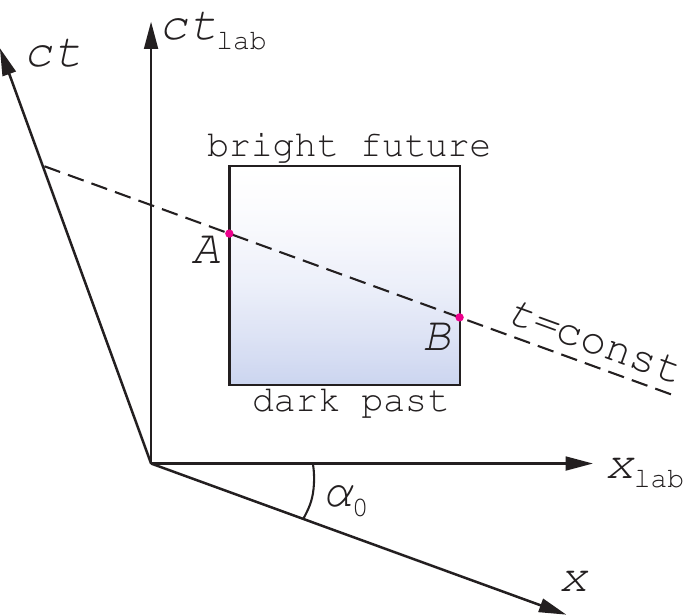}
\caption{Transformation of the coordinate system under Lorentz transformations, where $\alpha_0=\arctan(|V_{0x}|/c)$ and $V_{0x}<0$. In the lab frame $t_\text{lab}(A)>t_\text{lab}(B)$, but in the frame $S'$ events $A$ and $B$ are simultaneous, $t(A)=t(B)$. Thus, during the boost, the past and future of the lab frame $S$ are mixed.}
\label{transform}
\end{figure}

With such a change of reference frame, the initial homogeneity of
the system along the $x$-axis is lost due to the relativity of simultaneity. Changing the reference frame leads to a rotation of the simultaneity planes in spacetime. A slice $t={\rm const}$ in the new frame appears as a slanted line in $(c
t_{\rm lab},x_{\rm lab})$: points with different $x_{\rm lab}$ are taken at different laboratory times (see Fig.~\ref{transform}). If the system evolves, the Lorentz boost inevitably mixes different temporal stages; hence, the initial homogeneity along $x$ is lost. 

As an illustration, in Fig.~\ref{Lorentz} we present the Lorentz-transformed fields and densities for a fixed time $t=\text{const}$. A strong inhomogeneity along the $x$ coordinate is clearly visible: small values of $x$ correspond to the "future" of the homogeneous $S$ frame, and large values of $x$ -- to the "past". The transformation is based on evaluating the PIC simulation data at space-time coordinates corresponding to a constant proper time in the boosted frame, $t_{\rm lab}(x_{\rm lab})=t/\Gamma_0 - |\beta_{0x}|x_{\rm lab}$, where $t={\rm const}$ is the time in $S'$. For each spatial point $x_{\rm lab}$, we identified the two nearest lab-frame simulation snapshots that bracket the corresponding $t_{\rm lab}$, and performed linear interpolation between them to obtain the field values at that position and time. The final fields were stored in 2D arrays corresponding to the spatial grid in the boosted frame.

\subsection{Main equations}

In the boosted frame, only the moving beam feels the magnetic field, which deflects beam particles into filaments. The motionless background plasma ($\Gamma=1$) simply screens the resulting electrostatic field.

In the boosted frame $S'$, Maxwell's equations take the following form:
\begin{equation}
\begin{split}
    &\frac{\partial E_x}{\partial x}+\frac{\partial E_y}{\partial y}=4\pi\rho, \\
    &\frac{\partial E_x}{\partial y}-\frac{\partial E_y}{\partial x}=\frac{1}{c}\frac{\partial B_z}{\partial t}, \\
    &\frac{\partial B_z}{\partial y}=\frac{4\pi}{c}j_{x}+\frac{1}{c}\frac{\partial E_x}{\partial t}, \\
    &\frac{\partial B_z}{\partial x}=-\frac{4\pi}{c}j_{y}-\frac{1}{c}\frac{\partial E_y}{\partial t}.
\end{split}
\end{equation}
Since $x=\Gamma_0\left(x_{\text{lab}}-V_{0x}t_\text{lab}\right)$ and $t=\Gamma_0(t_\text{lab}-V_{0x}x_\text{lab}/c^2)$, where $V_{x0}$ is the velocity of $S'$ relative to lab-frame $S$ in which the problem is homogeneous, and in the lab-frame $S$ nothing depends on $x_\text{lab}$, we can write $\partial f/\partial x_\text{lab}=0$ for any $f$. This gives $\partial f/\partial x=(V_{0x}/c^2)\partial f/\partial t$. Thus, longitudinal spatial gradients can be replaced by time derivatives multiplied by $V_{0x}/c^{2}$. The price for removing all longitudinal structures from the problem is that relativistic effects appear, since the replacement of the derivative couples two reference frames $S$ and $S'$. An attempt to give a completely independent explanation in the $S'$ frame would force us to take into account the unknown longitudinal gradients of all quantities that are given by future and past time slices in the laboratory reference frame $S$.

\begin{figure}
\centering
\includegraphics[width=0.49\textwidth]{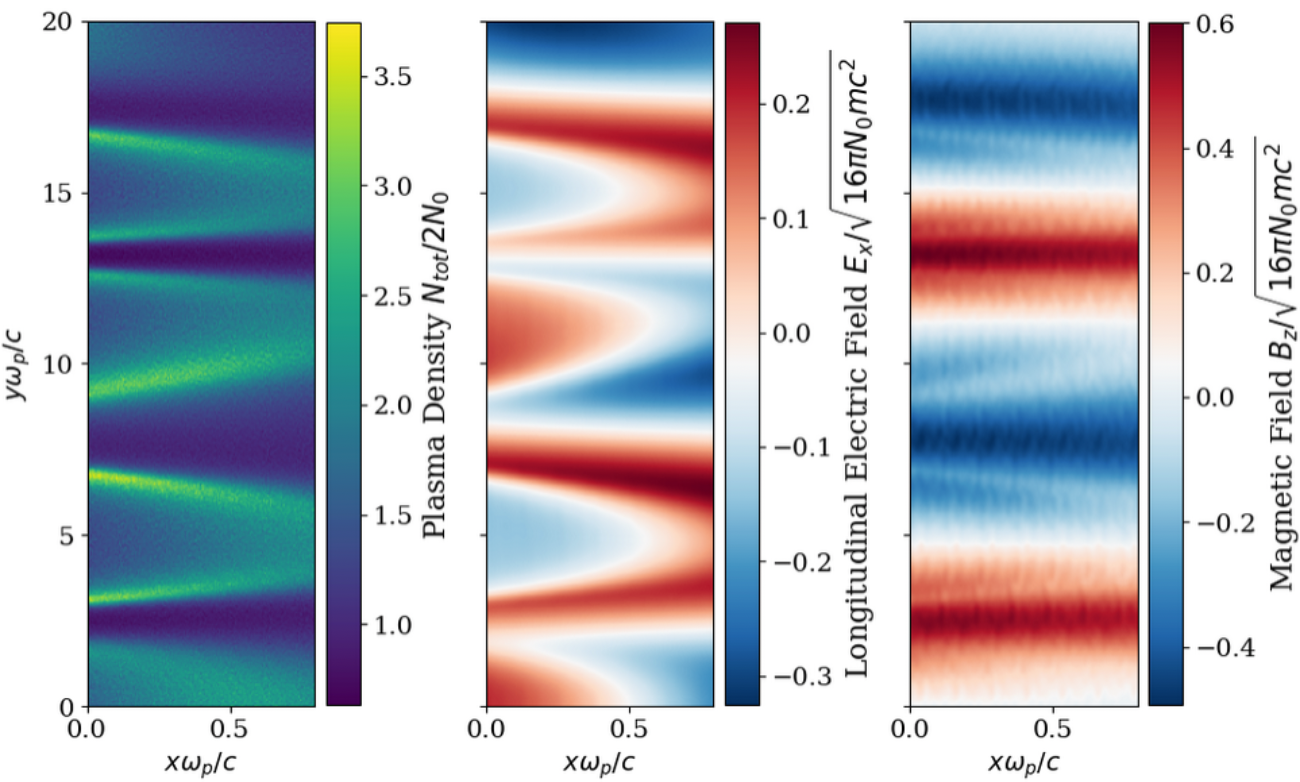}
\caption{Distribution of the background plasma density, longitudinal electric field $E_x$, and magnetic field $B_z$ in the boosted frame $S'$ at $t=\text{const}$ in this frame. The simulation parameters correspond to the reference run.}
\label{Lorentz}
\end{figure}

After substituting the $\partial/\partial x$ derivative, we obtain
\begin{equation}
\begin{split}
    &\frac{V_{0x}}{c^2}\frac{\partial E_x}{\partial t}=4\pi\rho-\frac{\partial E_y}{\partial y}, \\
    &\frac{1}{c}\frac{\partial}{\partial t}\left(B_z+\frac{V_{0x}}{c}E_y\right)=\frac{\partial E_x}{\partial y}, \\
    &\frac{1}{c}\frac{\partial E_x}{\partial t}=\frac{\partial B_z}{\partial y}-\frac{4\pi}{c}j_x, \\
    &\frac{1}{c}\frac{\partial}{\partial t}\left(E_y+\frac{V_{0x}}{c}B_z\right)=-\frac{4\pi}{c}j_y.
\end{split}
\end{equation}
It is easy to see that the second and last equations involve fields defined in the laboratory reference frame $E^\text{lab}_y=\Gamma_0[E_y+(V_{0x}/c)B_z]$ and $B^\text{lab}_z=\Gamma_0[B_z+(V_{0x}/c)E_y]$. This is a consequence of the fact that $j_y$ and $\partial/\partial y$ do not transform between frames of reference moving along $x$. Moreover, since in the laboratory reference frame everything was homogeneous along $x$, we obtain $\partial/\partial t=\Gamma_0(\partial/\partial t_\text{lab})$ and the Lorentz factor $\Gamma_0$ cancels out. That is, the second and last equations are simply Maxwell's equations in the lab frame $S$ (see equations~(\ref{maxwell})).

One can see that in such a description time intervals are reduced, $\text{d}t=\text{d}t_\text{lab}/\Gamma_0$, due to a combination of the temporary growth of structures and their drift along $x$ with the velocity $|V_{0x}|$. In other words, by following the evolution of fields at $x=\text{const}$, their drift through the section under consideration effectively accelerates their evolution.

We can transform the equations for the fields $E_y$ and $B_z$ to wave-type equations:
\begin{equation}
\begin{split}
\square^2B_z=\frac{4\pi}{c}\left(\frac{V_{0x}}{c^2}\frac{\partial j_y}{\partial t}-\frac{\partial j_x}{\partial y}\right), \\
\square^2E_y=-\frac{4\pi}{c}\left(\frac{1}{c}\frac{\partial j_y}{\partial t}-c\frac{\partial\rho}{\partial y}\right).
\end{split}
\end{equation}
These equations contain the d'Alembert operator
\begin{equation}
\square^2 = \left(\frac{1}{c\Gamma_0}\right)^2\frac{\partial^2}{\partial t^2}-\frac{\partial^2}{\partial y^2}.
\end{equation}
Considering only the left-hand side of the equations, the solution for a plane wave yields $\omega=\Gamma_0 k$. In fact, this means that across the filaments the phase velocity of the waves is enormous and equals $v_\text{ph}=c\Gamma_0\gg c$. This does not violate causality, but is simply due to the additional relativistic drift of the entire structure along $x$.

The estimation~(\ref{elec}) also works in the reference frame $S'$. As in $S$ it has an accuracy of order $\mathcal{O}(\Gamma_0^{-2})$ and gives a good estimation for $\partial E_x/\partial y$. However, we cannot use this estimation for the fields $B_z$ and $E_y$.
From Maxwell's equations, we can obtain
\begin{equation}\label{magneq}
\begin{split}
\frac{\partial B_z}{\partial t}=c\Gamma_0^2\left(\frac{\partial E_x}{\partial y}+\frac{4\pi V_{0x}}{c^2}j_y\right), \\ \frac{\partial E_y}{\partial t}=c\Gamma_0^2\left(\frac{V_{0x}}{c}\frac{\partial E_x}{\partial y}+\frac{4\pi}{c}j_y\right).
\end{split}
\end{equation}
These equations contain a factor $\Gamma_0^2$ which makes the error of order $\mathcal{O}(1)$.

After replacing the convective derivative $\partial/\partial x$ with a time derivative, the continuity equation is
\begin{equation}\label{plasma}
\frac{1}{\Gamma_0}\frac{\partial N^\text{lab}_s}{\partial t}+\frac{\partial}{\partial y}N_s v_{sy}=0,
\end{equation}
where $N^\text{lab}_s=\Gamma_0N_s(1+V_{0x}v_{sx}/c^2)$. The first term $\Gamma_0 N_s$ in $N^\text{lab}_s$ gives the usual increase in the density due to the volume contraction: the same number of particles occupy the smaller volume $\mathcal{V}_\text{lab}=\mathcal{V}/\Gamma_0$. The second term $\sim V_{0x}v_{sx}/c^2$ is connected with the mass flux and the relativity of simultaneity. If we consider $t_\text{lab}=\text{const}$ in the frame $S$, it corresponds to the time difference ${\rm d}t=-(V_{0x}/c^2){\rm d}x$ on the ends of small volume $\mathcal{V}$ in the frame $S'$. But we measure $N_s$ at $t=\text{const}$, therefore, one must correct for the particles that enter/leave through the volume $\mathcal{V}$ during ${\rm d} t$ owing to the non-zero mass flux $N_sv_{sx}$. Together with the volume contraction, it gives ${\rm d}N_s=-\Gamma_0 N_sV_{0x}v_{sx}/c^2$.

Substituting the Lorentz–transformed
density $N_s^\text{lab}$, the continuity equation in the boosted frame $S$ takes
the form
\begin{equation}
  \label{eq:cont_general}
  \frac{\partial N_s}{\partial t}
  = -\left(1+\frac{V_{0x}v_{sx}}{c^{2}}\right)^{-1}
    \left[
      \frac{\partial}{\partial y}\!\left(N_s v_{sy}\right)
      +\frac{V_{0x}N_s}{c^{2}}\,
       \frac{\partial v_{sx}}{\partial t}
    \right].
\end{equation}
Equation~\eqref{eq:cont_general} shows that the density evolves for two
distinct reasons:

\begin{enumerate}
\item \textit{Transverse mass flux.}
      The first term inside the brackets is the familiar divergence
      $\partial(N_s v_{sy})/\partial y$.
      Beam particles cross the filament under the action of the
      magnetic force, whereas background plasma particles move in a transverse direction to
      compensate for the beam charge via the electric force.

\item \textit{Relativistic compression.}
      The second term and denominator represent purely relativistic corrections and depend both on the plasma velocity $v_{sx}$ and the plasma acceleration $\partial v_{sx}/\partial t$.
\end{enumerate}

For the relativistic beam, the longitudinal inertia
$m\Gamma_{\rm b}^{3}$ is enormous
($\Gamma_{\rm b}\sim 2\Gamma_{\rm b}^\text{lab}\Gamma_{\rm p}^\text{lab}$), so
$\partial v_{{\rm b}x}/\partial t\simeq 0$ and the second term in
Eq.~\eqref{eq:cont_general} is negligible.
However, the denominator
$1+V_{0x}v_{{\rm b}x}/c^{2}$ is nearly zero because
$v_{{\rm b}x}\simeq c$ and $V_{0x}\simeq-c$; this dramatically amplifies the effect of the transverse mass flux and allows the beam density to dominate inside the filament.

Background positrons and electrons experience the
inductive field $E_x$ in opposite directions, so in the boosted frame~$S'$ their longitudinal velocities evolve with opposite signs.
For definiteness, consider a filament that has already become positron–dominated:
$v_{+x}>0$ (the positrons lag behind the frame $S$) while
$v_{-x}<0$ (the electrons move ahead).
Consequently, the denominator in Eq.~\eqref{eq:cont_general}
decreases in time for positrons and
increases for electrons.
The density of positrons, therefore, grows much faster than the electron
density decreases, producing the strong charge asymmetry inside the filament (this is why the filaments in Fig.~\ref{Lorentz} are so strong).

To describe the motion of plasma, the Euler equation is used,
\begin{equation}\label{eom}
\begin{split}
 \left(1+\frac{v_{sx} v_{0x}}{c^2}\right)\frac{\partial \mathbf{v}_s}{\partial t}&+v_{sy}\frac{\partial \mathbf{v}_s}{\partial y}= \\
 &=\frac{q_s}{\Gamma_s m}\left[\mathbf{E}+\frac{1}{c}(\mathbf{v}_s\times\mathbf{B})-\frac{\mathbf{v}_s(\mathbf{v}_s\cdot\mathbf{E})}{c^2}\right].
\end{split}
\end{equation}

\subsection{Cavity formation in the reference frame $S'$}

\begin{figure*}
\centering
\begin{minipage}[t]{0.47\linewidth}
    \centering
    \includegraphics[width=\linewidth]{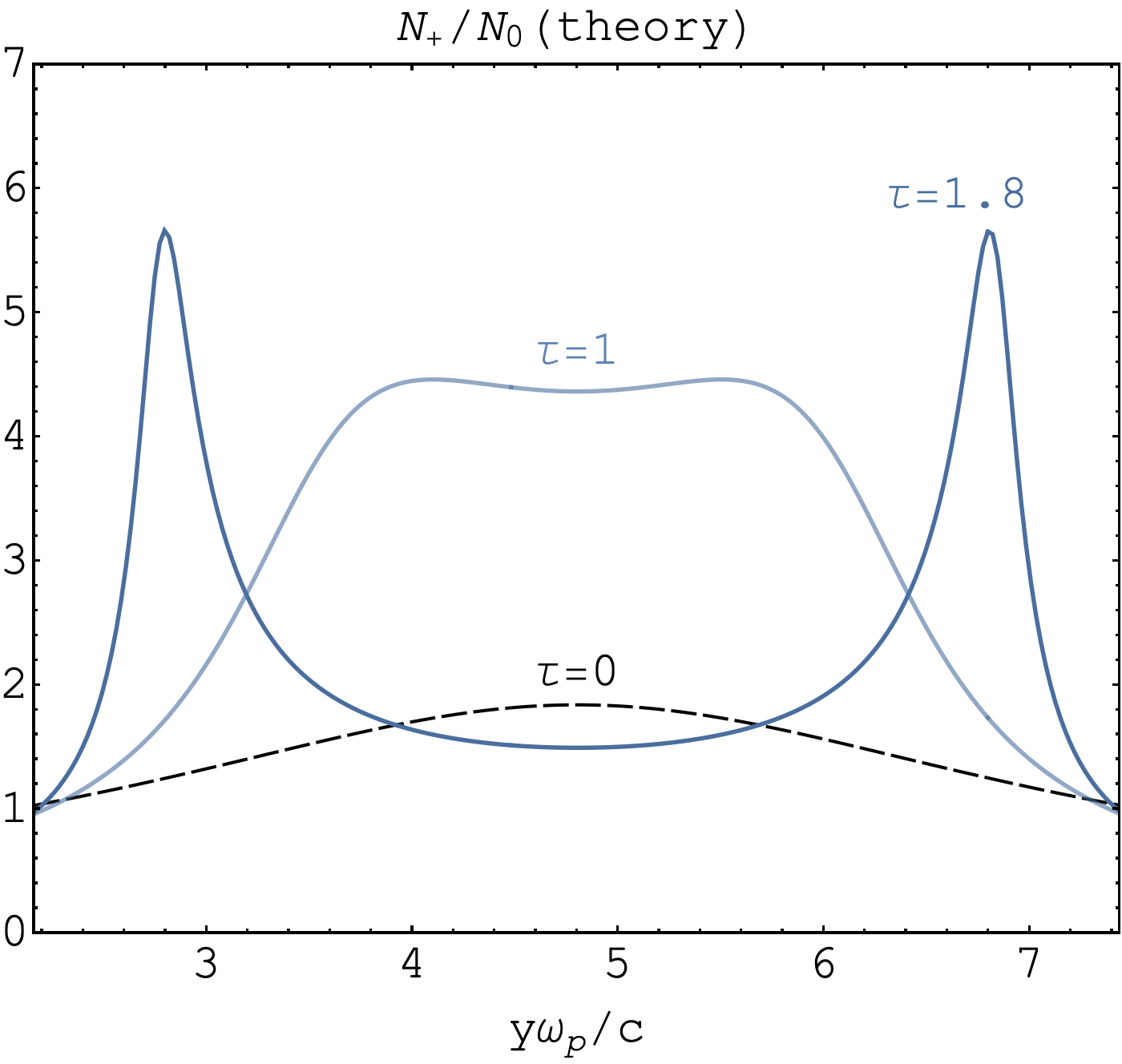}
\end{minipage}\hfill
\begin{minipage}[t]{0.5\linewidth}
    \centering
    \includegraphics[width=\linewidth]{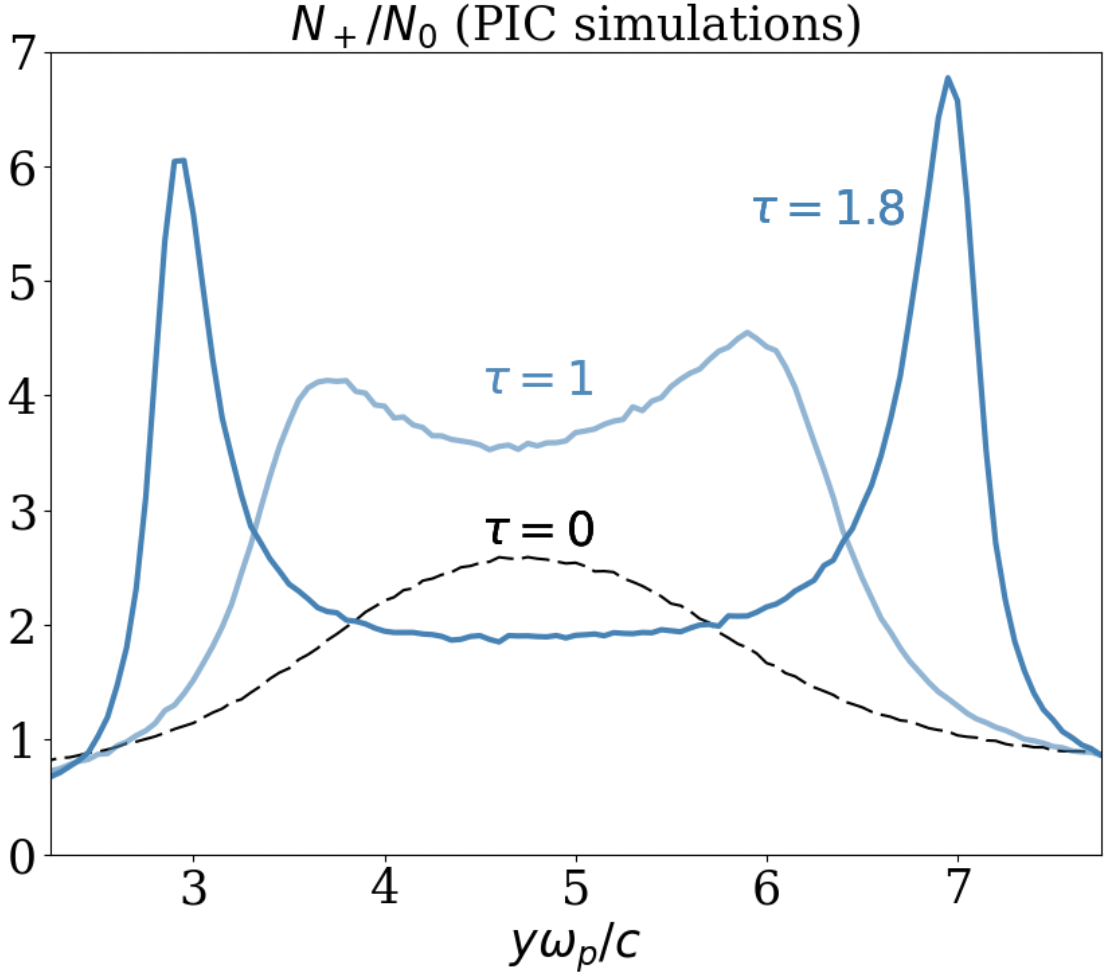}
\end{minipage}
\caption{Evolution of the positron density $n_+(\eta)$ in the initially co-moving frame $S'$: left — simple model; right — PIC simulations.}
\label{fig:Sprime-combined}
\end{figure*}

\begin{figure*}
\centering
\begin{minipage}[t]{0.45\linewidth}
    \centering
    \includegraphics[width=\linewidth]{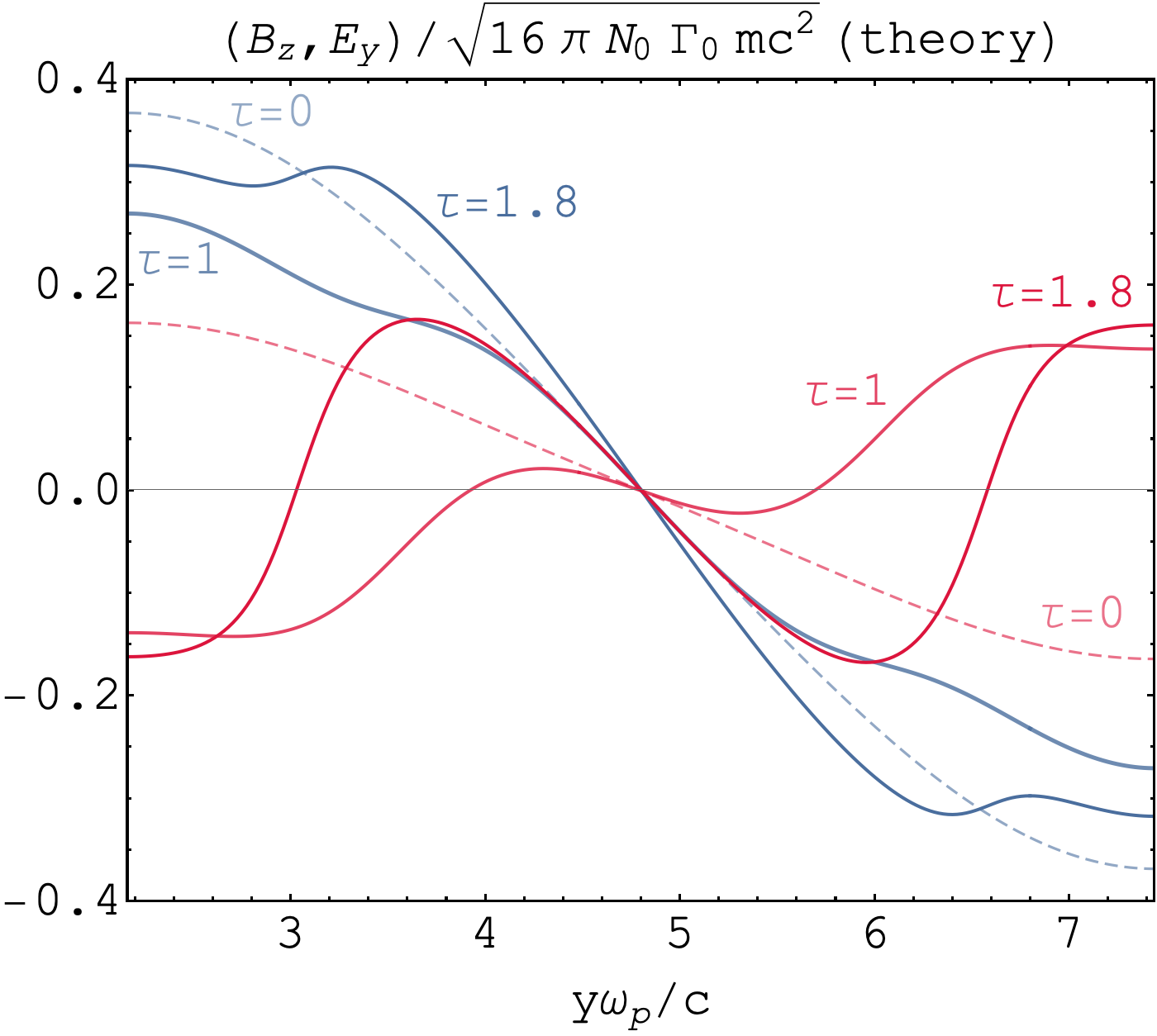}
\end{minipage}\hfill
\begin{minipage}[t]{0.5\linewidth}
    \centering
    \includegraphics[width=\linewidth]{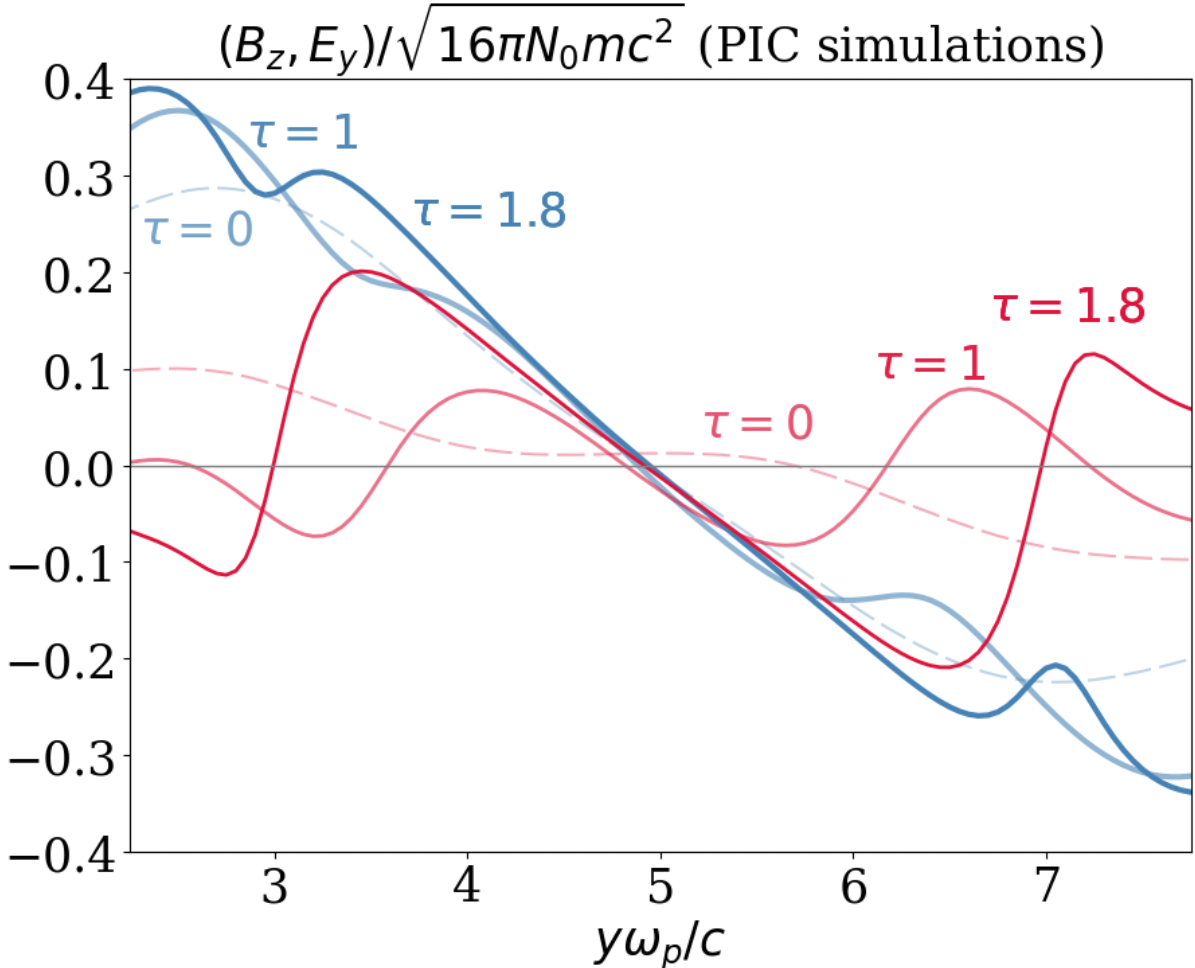}
\end{minipage}
\caption{Evolution of the fields $b_z$ and $\varepsilon_y$ in the initially co-moving frame $S'$: left — simple model; right — PIC simulations.}
\label{fig:Fprime-combined}
\end{figure*}

We again consider a plasma positron-dominated filament, assuming that the initial densities of the beam and background plasma were approximately equal in the $S'$ frame. Initially, the background plasma is motionless and does not create a current. Thus, during the linear stage of the Weibel instability, the beam particles in the filament set the total current and magnetic field. The increase in the magnetic field $\partial |B_z|/\partial t> 0$ induces a longitudinal electric field $E_x$ inside the filament, which decelerates the beam electrons and accelerates the plasma positrons in the beam direction. When the plasma positrons reach a certain velocity, relativistic effects begin to dominate and significantly increase the positron density due to their resulting lag relative to the considered reference frame. The resulting sharp increase in the positron current leads to the fact that in the center of the filament, the total current density begins to decrease rather than increase, $\partial j_{\text{tot}}/\partial t<0$. This reduces the magnetic field in the center, which in turn begins to decrease the field $E_x$ there. According to Lenz's rule, a decrease in the field $E_x$ slows down the decrease in $B_z$. But since the positrons are still slow at the edges of the filament, $E_x$ there is practically the same. As a result, the function $E_x(y)$ forms a double-humped profile: it has two lateral maxima with a rapidly decreasing minimum between them.

Since at any point of the filament $B_z^2-E_y^2\geq 0$, the positrons accelerated to relativistic velocities in the filament center begin to move from the filament center to the edges, since when moving in the same direction as the beam, the Lorentz force $(e/c)v_xB_z$ is defocusing for them and at a sufficiently high velocity $v_x$ it exceeds the holding electric force $eE_y$.
As soon as $E_x$ changes sign in the filament center, the plasma begins to slow down here and the Lorentz compression disappears, reducing the positron density to normal values comparable with the density of the beam electrons. But in the region where $E_x>0$ (two maxima), the plasma continues to accelerate and its density increases due to relativistic compression. Therefore, there are two density peaks that gradually move to the edges. They move to the edges for two reasons: the Lorentz force displaces particles from the center, without changing $v_{x}$ much, and thus preserving the relativistic compression. Also, the longitudinal field $E_x$ itself gradually accelerates new portions of positrons away from the center of the filament. The maxima of the electric field $E_x$ move synchronously with the maxima of the positron density: the field maintains the acceleration of positrons just before the density hump, and the density hump itself kills this field due to the decrease in the total current.

From the equations of Ampère and Gauss, we can obtain
\begin{equation}
\frac{\partial E_y}{\partial y}=\frac{4\pi}{c}\left(\rho c+j_x\frac{V_{0x}}{c}\right)-\frac{V_{0x}}{c}\frac{\partial B_z}{\partial y}.
\end{equation}
By writing expressions for the current and charge densities, taking into account only the dominant contributions, and using the approximation $V_{0x}\approx -c$, we obtain
\begin{equation}
\frac{\partial E_y}{\partial y}\approx 4\pi N_{+}\left(1 -\frac{v_{+x}}{c}\right)+\frac{\partial B_z}{\partial y}.
\end{equation}
From this, we conclude that if at the center of the filament the plasma density decreases $N_+\rightarrow0$ or the motion becomes relativistic $v_{+x}\rightarrow c$, then the slope of the electric field is equal to the slope of the magnetic field (see Fig.~\ref{fig:Fprime-combined}). Since both fields are zero at $y=0$ (symmetry of the system), then $E_y\rightarrow B_z$ is where $N_+\rightarrow 0$ or $v_{+x}\rightarrow c$ (in this case always $|E_y|<|B_z|$). It is easy to verify that due to the Lorentz transformations,
\begin{equation}
\begin{split}
    E_y^{\text{lab}}=\Gamma_0\left(E_y+\frac{v_{0x}}{c}B_z\right), \\
    B_z^{\text{lab}}=\Gamma_0\left(B_z+\frac{v_{0x}}{c}E_y\right),
\end{split}
\end{equation}
in the lab frame $S$, where both beams were moving towards each other, $E_y$ and $B_z$ practically cancel each other out ($v_{0x}<0$). Thus, inside a cavity, we obtain small magnetic and electric fields.

\section{Numerical setup}\label{sec:numerical}

\subsection{Homogeneous simulation}

\begin{table*}
    \centering
    \begin{tabular}{|c| c c c c c c | c c c c |}
        \hline
         & $n_{p}$ & $\Gamma_0$ & $T_p/m c^2$ & $n_b$ & $\Gamma_{b0}$ & $T_b/m c^2$ & $\omega_p/ck$ & $\Gamma_w$ & $\gamma_w/\omega_p$ & Type \\
        \hline
        run 1 & 5 & 10 & $10^{-3}$ & 0.03 & 3 & 10 & 1.3 & 7 & 0.088 & MS \\
        \hline
        run 2 & 5 & 10 & $10^{-3}$ & 0.01 & 3 & 10 & 1.9 & 9 & 0.032 & PS \\
        \hline
        run 3 & 5 & 10 & $10^{-3}$ & 0.03 & 30 & 10 & 0.9 & 7.5 & 0.087 & MS \\
        \hline
        run 4 & 3.5 & 7 & $10^{-2}$ & 0.03 & 3 & 10 & 1.3 & 6 & 0.059 & PS \\
        \hline
        run 5 & 7.5 & 15 & $10^{-3}$ & 0.05 & 3 & 10 & 1.4 & 8 & 0.13 & MS \\
    \end{tabular}
    \caption{Results from homogeneous PIC simulations. Here MS denotes "magnetic sandwich" and PS -- "plasma sandwich".}
    \label{tab1}
\end{table*}

The spatial and temporal
resolutions of the simulation are $\Delta x=0.05(c/\omega_p)$ and $\Delta t=0.025\omega_p^{-1}$. For noise suppression, we used 128 particles per cell, a binomial current filter (10 passes on the $x$-coordinate and 5 passes on the $y$-coordinate at each time step), a 5-points stencil for the particle shape function, and a 4th-order Maxwell solver "M4". The coefficients for the extended stencil in Faraday's equation are (\citealt{Lu2020})
\begin{equation}
\begin{split}
    &\beta_{xy}=\beta_{yx}=\frac{(c\Delta t/\Delta x)^2}{12}, \quad \delta_x=\delta_y=\beta_{xy}-\frac{1}{12}, \\
    &\alpha_x=\alpha_y=1-2\beta_{xy}-3\delta_x.
\end{split}
\end{equation}
For particles we used the relativistic Higuera-Cary pusher (\citealt{HigueraCary2017}).

\subsection{Shock simulation}

For the shock simulation, we used a setup similar to \cite{Groselj2024}. The shock was generated by reflecting a cold upstream flow (with Lorentz factor $\Gamma_0=10$) off the left boundary, while fresh upstream plasma was continuously injected from the right (e.g. \citealt{Spitkovsky2008}, \citealt{Sironi2013}). The reference run employed a 2D domain of $4096\times512(c/\omega_p)$, resolved with a cell size of $\Delta x = 0.1 (c/\omega_p)$ and a timestep of $\Delta t = 0.05 \omega_p^{-1}$. Each cell contained 16 particles per species. To mitigate numerical artifacts, electric currents were low-pass filtered at each step (10 passes on $x$-coordinate and 5 passes on $y$-coordinate), and electromagnetic fields were advanced using a modified Blinne stencil to suppress numerical Cherenkov instability \citep{Blinne2018}. The Maxwell solver "M4" \citep{Lu2020} leads to similar results.

The upstream flow Lorentz factor was initialized with a gradual increase from $\Gamma=1$ at the reflecting boundary to the far-upstream value $\Gamma_0=10$ over a transition layer of $200(c/\omega_p)$. This setup ensured a smooth shock formation and reduced artificial reflections (numerical precursor) from the left boundary.

\section{Plasma magnetization criterion}\label{app:magnetization}

The plasma Larmor radius in the Weibel frame is
\begin{equation}
    r_{L|w}=\frac{\Gamma_{0|w}mv_{0|w}c}{e B_{z|w}}=\frac{\Gamma_0\Gamma_w^2mc}{e B_z}\left(v_{0}-c\sqrt{1-\frac{1}{\Gamma^2_w}}\right),
\end{equation}
where we used $\Gamma_{0|w}=\Gamma_0\Gamma_w(1-v_{0x}V_{wx}/c^2)$, $v_{0x|w}=(v_{0x}-V_{wx})/(1-v_{0x}V_{wx}/c^2)$ and $B_{z|w}=B_{z}/\Gamma_w$. Since the field is zero at the center of the filament and is maximum at the edges, we take $ B_z=B_{z}^{\rm max}/2$ as the characteristic magnetic field.

The magnetization criterion is $r_{L|w}<\pi/2k$. Expressing the magnetic field amplitude in the dimensionless form, we obtain
\begin{equation}\label{d2}
    b_{\rm max}>\frac{4}{\pi}\left(\frac{ck}{\omega_p}\right)\Gamma_0\Gamma_w^2\left(\sqrt{1-\frac{1}{\Gamma_0^2}}-\sqrt{1-\frac{1}{\Gamma_w^2}}\right).
\end{equation}
Now we can understand whether the plasma is magnetized before the cavity is formed or not. The magnetic field corresponding to $\Gamma_+=\Gamma_w$ is determined by~(\ref{can}). Substituting it in~(\ref{d2}), we obtain
\begin{equation}
    \Gamma_0-\Gamma_w>\frac{4}{\pi}\Gamma_0\Gamma_w^2\left(\sqrt{1-\frac{1}{\Gamma_0^2}}-\sqrt{1-\frac{1}{\Gamma_w^2}}\right).
\end{equation}
In the ultrarelativistic limit, when $\Gamma_0\gg1$ and $\Gamma_w\gg 1$, one can obtain that plasma is magnetized when $\Gamma_w$ is smaller than $60\%$ of $\Gamma_0$. According to simulations (see Table~\ref{tab1}), the plasma is magnetized before the cavity formation if
\begin{equation}
    \Gamma_w\lesssim 0.8\Gamma_0,
\end{equation}
which is close to theoretical predictions.

\bsp	
\label{lastpage}
\end{document}